\documentclass[5p]{my_elsarticle}
\pdfoutput=1
\usepackage{lscape} 

\usepackage{relsize}
\usepackage{etoolbox}
\newtoggle{submission}
\toggletrue{submission}
\newtoggle{TwoColumnFormulas}
\togglefalse{TwoColumnFormulas}
\biboptions{longnamesfirst,square,sort,compress}

\usepackage[english]{babel}
\usepackage[utf8]{inputenc}
\usepackage[T1]{fontenc}   
\usepackage{graphicx}
\usepackage{enumerate}

\usepackage{color,amssymb,amsmath}
\usepackage{array}
\usepackage{upgreek}
\usepackage{txfonts}
\usepackage{empheq}
\usepackage{mathtools}
\usepackage{mathrsfs} 
\DeclareMathAlphabet{\mathscrbf}{OMS}{mdugm}{b}{n} 
\usepackage{frcursive}
\usepackage{comment}
\usepackage{booktabs}
\usepackage{placeins}

\usepackage{tikz}

\usepackage{pgfplots}
\iftoggle{submission}{}{
  \usetikzlibrary{pgfplots.groupplots}
  \usetikzlibrary{calc}
  \pgfplotsset{compat=1.12}
  \usepgfplotslibrary{external}
  \tikzexternalize
}

\iftoggle{submission}{
  
}{

}

\newlength{\GraphWidth}
\newlength{\GraphHeight}
\newlength{\GraphHorizSep}
\newlength{\GraphVerticSep}
\newlength{\GraphVerticLegendSep}
\newlength{\HorizVerticLegendSep}
\makeatletter
\if@twocolumn
  \tikzset{every mark/.append style={scale=1.2,line width=0.3pt}}
  \pgfplotsset{legend image code/.code={\draw[mark repeat=3, mark phase=2,#1] plot coordinates {(0cm,0cm)(0.3cm,0cm)(0.6cm,0cm)};\draw[color=white] (0cm,-0.07cm) rectangle (0.6cm,0.07cm);}}
  \newenvironment{SmartFigure}{\begin{figure}}{\end{figure}}
  \newcommand{\MyLWone}{0.75pt}  
  \newcommand{\MyLWtwo}{1pt} 
\else
  \tikzset{every mark/.append style={line width=0.6pt}}
  \pgfplotsset{legend image code/.code={\draw[mark repeat=3, mark phase=2,#1] plot coordinates {(0cm,0cm)(0.5cm,0cm)(1.0cm,0cm)};\draw[color=white] (0cm,-0.1cm) rectangle (1.0cm,0.1cm);}}
  \newenvironment{SmartFigure}{\begin{figure*}}{\end{figure*}}
  \newcommand{\MyLWone}{1pt}  
  \newcommand{\MyLWtwo}{2pt} 
\fi
\pgfplotscreateplotcyclelist{MyCycleListOne}{{black,only marks,mark=star,mark options={scale=1.3},mark repeat={10}},{blue,densely dotted,line width = \MyLWone},{green,only marks,mark=o,mark repeat={10}},{red,loosely dotted,line width = \MyLWtwo},{black}}
\pgfplotscreateplotcyclelist{MyCycleListTwo}{{green,line width = \MyLWone},{black,loosely dotted,line width = \MyLWone},{blue,densely dotted,line width = \MyLWone},{red,loosely dotted,line width = \MyLWtwo},{black}}
\pgfplotscreateplotcyclelist{MyCycleListThree}{{black,dashed},{green,densely dashed,line width=0.75pt},{red,dash pattern=on 0.75pt off 0.75pt on 5pt off 0.75pt},{blue,densely dotted,line width=0.75pt},{black}}
\makeatother

\usepackage{cancel}

\allowdisplaybreaks[4]

\specialcomment{myproofs}
{\begingroup\color{blue}}{\endgroup}
\specialcomment{Tomodify}
{\begingroup\color{red}}{\endgroup}
\excludecomment{myproofs}

\usepackage{hyperref}
\hypersetup{
    unicode=true,          
    pdftoolbar=true,        
    pdfmenubar=true,        
    pdffitwindow=true,     
    pdfstartview={FitH},    
    pdftitle={A family of higher-order single layer plate models meeting \protect{$C^0_z$}-requirements for arbitrary laminates},    
    pdfauthor={Alexandre Loredo},     
    pdfcreator={Alexandre Loredo},   
    pdfproducer={Alexandre Loredo}, 
    pdfnewwindow=true,      
    colorlinks=true,       
    linkcolor=red,          
    citecolor=green,        
    filecolor=magenta,      
    urlcolor=cyan           
}

\usepackage[noabbrev]{cleveref}

\newcommand{\wfs}{\emph{warping functions}}
\newcommand{\Wfs}{\emph{Warping functions}}
\newcommand{\sfs}{\emph{stress functions}}
\newcommand{\tosdt}{ToSDT}
\newcommand{\tozzfour}{ToZZ}
\newcommand{\sizzfour}{SiZZ}
\newcommand{\hyzzfour}{HyZZ}

\newcommand{\hyspe}{HySpe}

\input cyracc.def

\journal{Composite Structures, accepted}

\usepackage[position=below,font=footnotesize,labelfont=bf,skip=10pt]{caption}
\begin{document}

\relscale{0.9}

\begin{frontmatter}
%
  \title{
  A family of higher-order single layer plate models meeting $C_z^0-$requirements for arbitrary laminates
  }
	\author[aff]{A.~Loredo\corref{cor1}}
	\ead{alexandre.loredo@u-bourgogne.fr}
	\cortext[cor1]{Corresponding author}
	\address[aff]{LEME, Universit\'e de Bourgogne--Franche-Comt\'e, France}
	\author[aff2]{M.~D'Ottavio}
	\author[aff2]{P.~Vidal}
	\author[aff2]{O.~Polit}
  \address[aff2]{LEME, UPL, Univ. Paris Nanterre, France}
	%
	%
	\begin{abstract}
In the framework of displacement-based equivalent single layer (ESL) plate theories for laminates, this paper presents a generic and automatic method to extend a basis higher-order shear deformation theory (polynomial, trigonometric, hyperbolic\dots) to a multilayer $C^0_z$ higher-order shear deformation theory. The key idea is to enhance the description of the cross-sectional warping: the odd high-order $C^1_z$ function of the basis model is replaced by one odd and one even high-order function and including the characteristic zig-zag behaviour by means of piecewise linear functions. In order to account for arbitrary lamination schemes, four such piecewise continuous functions are considered. The coefficients of these four warping functions are determined in such a manner that the interlaminar continuity as well as the homogeneity conditions at the plate's top and bottom surfaces are {\em a priori} exactly verified by the transverse shear stress field. These $C_z^0$ ESL models all have the same number of DOF as the original basis HSDT. Numerical assessments are presented by referring to a strong-form Navier-type solution for laminates with arbitrary stacking sequences as well for a sandwich plate. In all practically relevant configurations for which laminated plate models are usually applied, the results obtained in terms of deflection, fundamental frequency and local stress response show that the proposed zig-zag models give better results than the basis models they are issued from.

	\end{abstract}
\end{frontmatter}
%
%

\section{Introduction}
Among the numerous theories that have been developed for multilayered plates, those belonging to the Equivalent-Single Layer (ESL) family are of practical interest due to their relatively small number of unknowns, that is independent of the number of layers. Within this ESL class, the classical lamination theory (CLT) which has been proposed first, does not take into account the transverse shear behaviour and is, therefore, accurate only for thin plates, for which the transverse shear deformation can be neglected. First order shear-deformation theories (FSDT) have then been proposed to overcome this problem upon retaining a transverse shear deformation that is constant throughout the plate's thickness. Its accuracy with regard to the plate's gross response (transverse deflection, low vibration frequencies) results nevertheless dependent on shear correction factors. Higher-order shear deformation theories (HSDT) have been subsequently proposed in order to avoid the need of these problem-dependent shear correction factors. This is accomplished upon describing the through-thickness behaviour of the in-plane displacement field by means of functions of order greater than one, which thus introduces an enhanced description of the transverse shear deformation. The most well-known HSDT is the Vlasov--Levinson--Reddy's third order theory (\tosdt{}). Polynomial functions are not the unique way to enrich the kinematic field, a wide variety of functions have been used, in particular trigonometric, hyperbolic, and exponential functions, as summarized in recent review papers~\cite{Abrate2017,Sayyad2017}. 
\par
In all the theories cited above, the transverse shear deformation is included into the kinematic field by means of functions of class $C^1$ along the thickness direction $z$. This leads to continuous transverse shear strains and hence to a discontinuous transverse shear stress field, which violates equilibrium conditions within multilayered structures. Authors have thus proposed to use piecewise continuous, differentiable functions of $z$, often referred to as zig-zag functions. These functions are commonly constructed in such a manner that appropriate jumps of their derivatives at the interfaces restore the transverse stress continuity. Such Zig-Zag (ZZ) theories have appeared around the half of the twentieth century with Ambartsumyan \cite{Ambartsumyan1958, Ambartsumian1970}, and Osternik and Barg~\cite{Osternik1964}, and have been continuously receiving attention until now. An earlier paper that belongs to this category is due to Lekhnitskii~\cite{Lekhnitskii1935}, but it was limited to the study of beams. The approach by Lekhnitskii has been extended to plates 50 years later by Ren~\cite{Ren1986a}. While the approaches by Ambartsumyan and Lekhnitskii rely on the exact verification of the constitutive equation connecting the transverse shear stress and the kinematic fields, Murakami \cite{Murakami1986} formulated a ZZ theory by postulating these two fields in an independent manner thanks to Reissner's mixed variational theorem (RMVT) dedicated to multilayered plates \cite{Reissner1984}. For more recent developments of RMVT-based ZZ theories, the interested reader may refer to papers by Carrera \cite{Carrera1996}, Demasi \cite{Demasi2009-p4} and Tessler \cite{Tessler2015,Iurlaro2015}. Murakami's zig-zag function (MZZF) has been also extensively applied to classical displacement-based variable kinematics approaches, see, e.g., \cite{Carrera2004,Demasi2013}. Within a comprehensive discussion about ZZ theories, Groh and Weaver have recently proposed a mixed ZZ theory based on Hellinger-Reissner's principle \cite{Groh-Weaver-ZZ-2015}.

Among these several approaches, more details of which can be found in the review papers~\cite{Carrera2003,Kreja2011,Khandan2012}, only a subset of these ZZ-theories are able to satisfy the appropriate interlaminar continuity (IC) of both, the displacement and the transverse shear stress fields. These two requirements (ZZ and IC) have been summarized by the acronym $C_z^0$--requirements~\cite{Carrera1995}. In the following, we shall limit our attention to those ZZ theories that satisfy {\em exactly} the $C_z^0-$requirements. A more detailed examination is next proposed of some pioneering works, with the aim of establishing the background and highlighting the differences with the family of models proposed in this paper. 
\begin{itemize}
	\item Ambartsumyan's approach is based on the early paper \cite{Ambartsumyan1958}, in which it is assumed that the transverse shear stress vary along $z$ according to a quadratic parabola with nil values at the outer surfaces. On page 20 of the later book \cite{Ambartsumian1970}, this assumption is formally expressed by the expressions $\tau_{xz}=f_1(z)\varphi(x,y)$ and $\tau_{yz}=f_2(z)\psi(x,y)$. It leads to 4 kinematic functions, but only two of them are independent, based on the primitives of $f_1(z)$ and $f_2(z)$. Conceived for orthotropic shells, the theory presented in the 1958 paper is not yet a ZZ theory, but in the 1970 book~\cite{Ambartsumian1970}, page 75, an extension to symmetric multilayered orthotropic plates is given, which exhibits 4 zig-zag kinematic functions issued from two modified functions $f_1(z)$ and $f_2(z)$. This extension is perhaps due to Osternik and Barg~\cite{Osternik1964}, which is cited in the 1970 Ambartsumyan's book (see also Carrera's review paper~\cite{Carrera2003}). 
	\item In 1969, Whitney extends Ambrartsumyan's theory to anisotropic plates, more specifically to general symmetric laminates and orthotropic non-symmetric laminates~\cite{Whitney1969}. Following Ambartsumyan's approach, Whitney starts from an assumed transverse shear stress field and ends up with 4 kinematic functions that are expressed as the superimposition of a polynomial third order \tosdt{} function and a zig-zag linear functions, see equation 6 of \cite{Whitney1969}. 
  \item Sun and Whitney propose in 1973 a layerwise model, which is the starting point for deriving an ESL model upon eliminating the parameters of the $N-1$ upper layers \cite{Sun1973}. The resulting ESL model is equivalent to a first order zig-zag model with 4 kinematic functions. The link between such models has been discussed in detail in \cite{Loredo2013}. It has constant transverse shear stresses, which is the drawback of first-order models. 
  \item The 1986 paper by Ren \cite{Ren1986a} proposes 4 kinematic functions from \emph{a priori} given transverse shear stress functions and applies the resulting zig-zag model to cross-ply laminates. Four displacement unknowns are introduced to take into account the transverse shear behaviour, which yields a 7-parameter model which is difficult to compare with our 5-parameter models.
  \item Cho and Parmerter formulate a ZZ theory for symmetric \cite{Cho1992} and general \cite{Cho1993} orthotropic composite plates. Starting point are 2 kinematic functions which are the superimposition of a cubic polynomial and a linear zig-zag function expressed in terms of the Heaviside function. The coefficients are determined by enforcing the transverse stress continuity, which leads to the coupling of the $x-$ and $y-$directions in the kinematic field. It therefore appears that the theory is in fact based on 4 kinematic functions (see equation 4 of \cite{Cho1993}). In his historical review, Carrera demonstrated the equivalence of Cho and Parmerter's model with Ambartsumyan's model. 
  \item It is finally worth mentioning the ZZ theory fulfilling the $C^0_z$--requirements that is based on trigonometric functions and developed by Ossadzow and coworkers  \cite{Ossadzow1998, Ossadzow2003}. The construction of the kinematic zig-zag functions follows a similar path as proposed by Cho and Parmerter, but the trigonometric $\cos$ and $\sin$ functions replace the quadratic and cubic terms of the polynomial expansion, respectively.   
\end{itemize}
\par
A conforming finite element based on a trigonometric ZZ theory enhanced through a transverse normal strain has been developed for laminated plates~\cite{Vidal2013}. References~\cite{Wang2015,Suganyadevi2016} extend the procedure of Cho and Parmerter to a wider family including polynomial, trigonometric, exponential and hyperbolic functions. However, the authors only consider two kinematic functions, which reduces the applicability of their models to cross-ply laminates. 
\par
In \cite{Loredo2014,Loredo2016}, corresponding polynomial and trigonometric $C^0_z$ zig-zag models have been constructed from basis polynomial and trigonometric HSDT by using four functions, which shall be hereafter referred to as \wfs{}. While in these works the \wfs{} were obtained from transverse stress fields obtained from 3D solutions or from equilibrium equations, the present paper presents a general procedure for extending a basis higher-order shear deformation theory (bHSDT) to a multilayer higher-order shear deformation theory (mHSDT) that meets the $C_z^0-$requirements. This extension consists in the construction of four $C^0_z$ \wfs{} starting from the native functions that characterize the basis theory: it can be applied to any couple of odd and even functions and has no limitation concerning the lamination scheme. 

The paper is organized as follows. Section \ref{sec:def} introduces the notation and points out the properties that the four \wfs{} are required to fulfil. The extension of a bHSDT up to an mHSDT fulfilling all $C^0_z$--requirements is described in Section \ref{sec:ext}. Three different basis models are exemplarily considered, which pertain respectively to the polynomial, trigonometric and hyperbolic type. It is also shown that the \wfs{} are components of a second-order tensor, hence being covariant with rotations about the $z-$axis. Section \ref{sec:results} reports the numerical evaluations: the $C^0_z$ \wfs{} effectively increase the accuracy of the basis (non zig-zag) model and this enhancement is quite insensitive with respect to the type of functions used for the model. A discussion is proposed in Section~\ref{sec:discussion} in order to substantiate the limitations of conventional ZZ models with respect to particularly ``constrained'' configurations with very low number of layers and length-to-thickness ratios: in these cases, accuracy may only be assured by resorting to \wfs{} that contain more layer-specific information, just as LayerWise models do. The main conclusions are finally summarised in Section~\ref{sec:conclusion}.
\section{Definitions and general properties\label{sec:def}}
This paper deals with a generic method to extend a basis higher-order shear deformation theory (bHSDT) to a multilayer higher-order shear deformation theory (mHSDT). This Section introduces the notation employed for identifying the various plate theories along with the fundamental properties that the underlying approximating functions are required to satisfy. 

\subsection{The basis theories}
We consider a basis high-order shear deformation theory (bHSDT) for which the kinematic field can be written in the following general form:
\begin{subequations}
\label{eq:depl}
\begin{empheq}[left=\empheqlbrace]{align}
  u_{\alpha}(z) &= u^0_{\alpha} - zw^0_{,\alpha} + \phi(z)\gamma^{0^-}_{\alpha3} \\
  u_{3}(z) &= w^0
\end{empheq}
\end{subequations}
where $u^0_{\alpha}$ are the membrane displacements at $z=0$, $w^0$ is the deflection at $z=0$, $\gamma^{0^-}_{\alpha3}$ are the transverse engineering strains at $z=0^-$, and  $\phi(z)$ is a $C^1$ odd function. The choice of the $0^-$ coordinate is a convention useful to avoid undetermined shear strains if an interface lies at $z=0$. The $z=0$ plane is assumed to be the middle plane of the plate, the lower and upper faces are respectively located at $z=-h/2$ and $z=h/2$.
\par
Written in the form reported in equation~\eqref{eq:depl}, the function $\phi(z)$ must verify $\phi(0)=0$ and $\phi'(0^-)=\phi'(0)=1$ to give sense to the notations, and $\phi'(\pm h/2)=0$ to enforce null transverse shear stresses at the top and bottom of the plate. Due to the $C^1$ property of $\phi(z)$, such theories do not have particular abilities to deal with multilayered plates. Indeed, the continuity of $\phi'(z)$ induces discontinuities of the transverse shear stresses $\sigma_{\alpha3}$ at the interfaces. Table~\ref{tab:bHSDTFunctions} summarizes the functions $\phi(z)$, along with the reference author and the type of the approximation, that will be extended to a ZZ model in Section~\ref{sec:ext}. 
\subsection{The multilayer HSDT}
A multilayer theory is a plate theory which is dedicated to composite plates upon fulfilling the $C^0_z$--requirements. The generic expression for the kinematics of a multilayer HSDT (mHSDT) is of the form:
\begin{subequations}
\label{eq:deplm}
\begin{empheq}[left=\empheqlbrace]{align}
    u_\alpha(z) & =  u_\alpha^0 - z w^0_{,\alpha} + \varphi_{\alpha\beta}(z)\gamma^{0^-}_{\beta3} \label{eq:in_plane_depl} \\ 
    u_3(z) & =  w^0 \label{eq:normal_depl}
\end{empheq}\label{eq:mHSDT}
\end{subequations}
where $\varphi_{\alpha\beta}(z)$ are four piecewise $C^1$ functions, sometimes called \wfs{}, that are requested to fulfil specific properties, as it will be discussed below. Among these properties, specific jump values need to be prescribed to their derivatives for enforcing continuity of transverse stresses at the interfaces. In order to construct an mHSDT that is applicable to arbitrary laminates, it is important to consider {\em four} functions $\varphi_{\alpha\beta}(z)$, and hence to retain the coupling between $\gamma^{0^-}_{13}$ (resp. $\gamma^{0^-}_{23}$) and $u_2$ (resp. $u_1$). In fact, models written with only two functions, i.e., with $\varphi_{12}(z)=\varphi_{21}(z)=0$, are only applicable to cross-ply laminates.
\begin{table}[htbp]
	\centering\small
		\begin{tabular}{ccc}
	\hline \hline
        Model & Type & $\phi(z)$                                            
      \\ \hline 
        Reddy\cite{Reddy1984} & Polynomial & $z-\dfrac{4z^3}{3h^2}$              
      \\ \noalign{\smallskip}
        Touratier\cite{Touratier1991} & Trigonometric & $\dfrac{h}{\pi}\sin\left(\pi\dfrac{z}{h}\right)$ 
      \\ \noalign{\smallskip}
        Soldatos\cite{Soldatos1992} & Hyperbolic & $\dfrac{\cosh\left(\frac{k}{2}\right)z-\frac{h}{k}\sinh\left(k\frac{z}{h}\right)}{\cosh\left(\frac{k}{2}\right)-1}$
      \\ \noalign{\smallskip}
	\hline \hline
		\end{tabular}
	\caption{Original bHSDT $\phi(z)$ functions. The parameter  $k$ of the hyperbolic function allows to generalise the original model with $k=1$ \cite{Soldatos1992}.}
	\label{tab:bHSDTFunctions}
\end{table}
\subsection{Required properties for the functions $\varphi_{\alpha\beta}(z)$}
Formula~\eqref{eq:in_plane_depl} shows that the piecewise functions $\varphi_{\alpha\beta}(z)$ must be continuous at each interface to respect the continuity of the in-plane displacements. Since $u_\alpha^0$ denotes the membrane displacements at $z=0$, the functions $\varphi_{\alpha\beta}$ must fulfil the following {\em homogeneity condition} 
\begin{equation}
\varphi_{\alpha\beta}(0)=0
\end{equation}
Figure~\ref{fig:Exemple1} illustrates the continuity and homogeneity conditions for a practical example (a $[-10/0/40]$ laminate, trigonometric mHSDT). 
\par
The compatible strain field defined by generic mHSDT kinematic field of Eq.~\eqref{eq:deplm} reads
\begin{subequations}
\label{eq:strains}
\begin{empheq}[left=\empheqlbrace]{align}
    \varepsilon_{\alpha\beta}(z) & =  \varepsilon^0_{\alpha\beta} - \tfrac{z}{2} w^0_{,\alpha\beta} + \tfrac{1}{2}(\varphi_{\alpha\gamma}(z)\gamma^{0^-}_{\gamma3,\beta}+\varphi_{\beta\gamma}(z)\gamma^{0^-}_{\gamma3,\alpha}) \label{eq:in_plane_strains} \\ 
    \varepsilon_{\alpha3}(z) & =  \tfrac{1}{2} \varphi'_{\alpha\beta}(z)\gamma^{0^-}_{\beta3} \label{eq:transverse_strains} \\
    \varepsilon_{33}(z) & =  0 \label{eq:normal_strain}
\end{empheq}
\end{subequations}
The transverse shear strains $\varepsilon_{\alpha3}(z)$ must be defined in each layer, but they also should be discontinuous at the interfaces for allowing the transverse shear stresses $\sigma_{\alpha3}(z)$ to be continuous in order to fulfil the equilibrium condition. Indicating by $\zeta_i$ the $z$-coordinate of the $i$th interface, with $i=1,2 \ldots N-1$, $\zeta_0=-h/2$ and $\zeta_N=h/2$, the functions $\varphi_{\alpha\beta}(z)$ are thus required to be piecewise $C^1$ over the intervals $]\zeta_{i-1}, \zeta_{i}[$. 
%
%
Furthermore, since $\gamma^{0^-}_{\beta3}$ represents the engineering shear strain at $z=0^-$, the derivatives of the functions $\varphi_{\alpha\beta}(z)$ are required to fulfil the following {\em homogeneity conditions}: 
\begin{equation}
\varphi'_{\alpha\beta}(0^-)=\delta_{\alpha\beta}
\end{equation}
Figure~\ref{fig:Exemple2} illustrates these conditions with the same practical example as before.

Due to the continuity of the $\varphi_{\alpha\beta}(z)$, in-plane strains $\varepsilon_{\alpha\beta}(z)$ are continuous. Following the classical plate approach, the normal stress $\sigma_{33}(z)$ is set to $0$, which leads to the use of reduced (in-plane) stiffnesses $Q_{\alpha\beta\gamma\delta}(z)$ in place of the stiffnesses $C_{\alpha\beta\gamma\delta}(z)$. It is further recalled that there is no physical reason for the in-plane stresses $\sigma_{\alpha\beta}(z)$ to be continuous at interfaces between adjacent layers with dissimilar stiffness coefficients.

\iftoggle{submission}{}{\tikzsetnextfilename{ICCS_Exemple1}}
\begin{SmartFigure}[htbp]
  \centering
  \iftoggle{submission}{
    \includegraphics{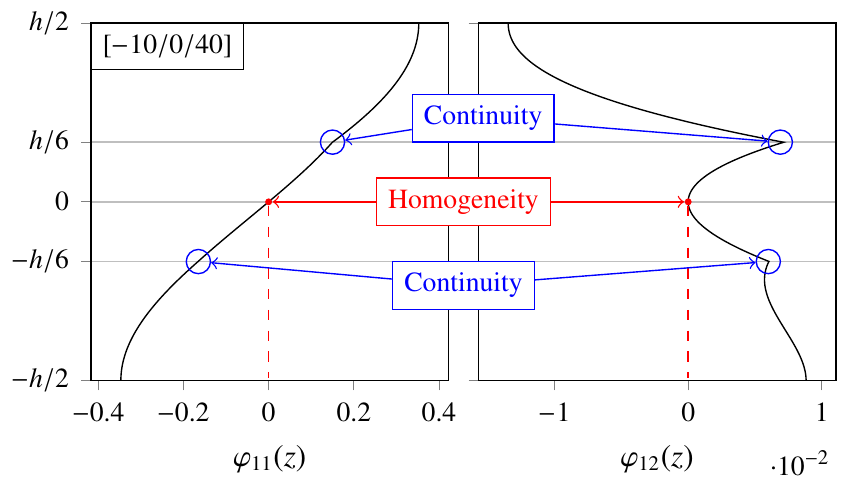}
  }{
    \input{Figures/ICCS_Exemple1.tex}
  }
  \caption{Illustration of the continuity and homogeneity conditions prescribed on the $\varphi_{\alpha\beta}(z)$ functions (a $[-10/0/40]$ laminate, trigonometric mHSDT). Only two of the four functions have been plotted.}%
  \label{fig:Exemple1}
\end{SmartFigure}

\iftoggle{submission}{}{\tikzsetnextfilename{ICCS_Exemple2}}
\begin{SmartFigure}[htbp]
  \centering
  \iftoggle{submission}{
    \includegraphics{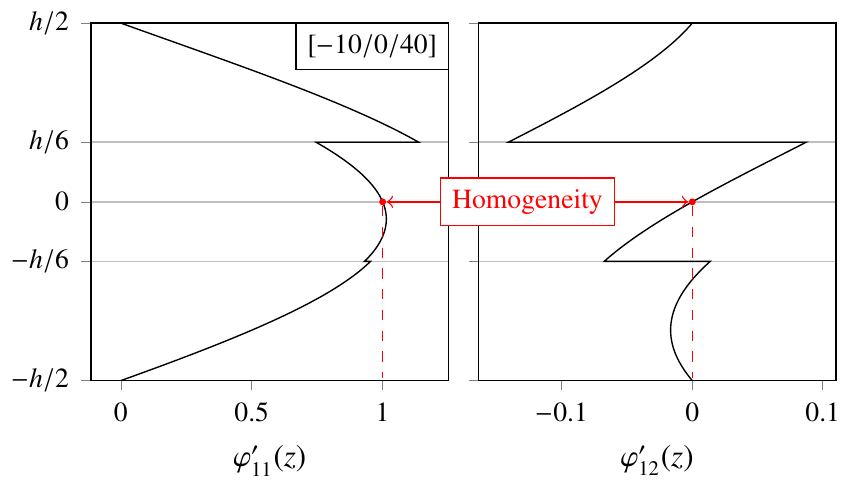}
  }{
    \input{Figures/ICCS_Exemple2.tex}
  }
  \caption{Illustration of the homogeneity conditions prescribed on the $\varphi'_{\alpha\beta}(z)$ functions (a $[-10/0/40]$ laminate, trigonometric mHSDT). Only two of the four functions have been plotted.}%
  \label{fig:Exemple2}
\end{SmartFigure}

\iftoggle{submission}{}{\tikzsetnextfilename{ICCS_Exemple3}}
\begin{SmartFigure}[htbp]
  \centering
  \iftoggle{submission}{
    \includegraphics{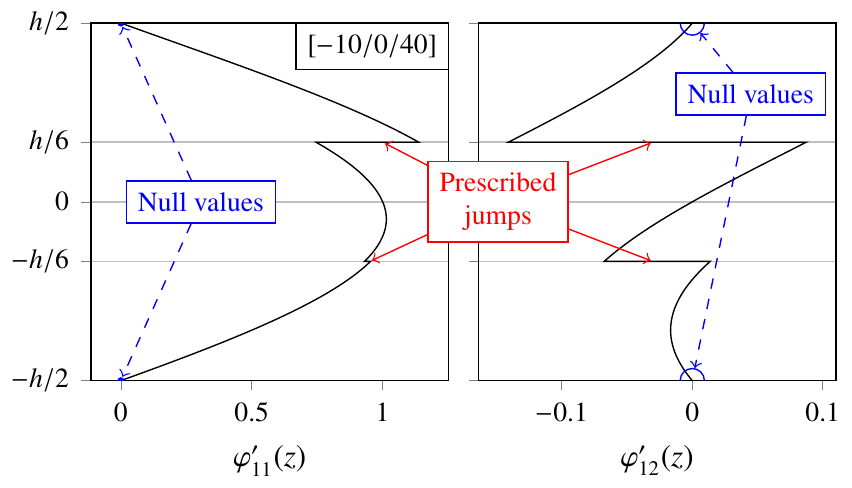}
  }{
    \input{Figures/ICCS_Exemple3.tex}
  }
  \caption{Illustration of the homogeneous bottom and top conditions, and the jump values prescribed on the $\varphi'_{\alpha\beta}(z)$ functions (a $[-10/0/40]$ laminate, trigonometric mHSDT). Only two of the four functions have been plotted.}%
  \label{fig:Exemple3}
\end{SmartFigure}
The constitutive equation defines the transverse shear stresses in terms of strains as follows:
\begin{equation}
  \sigma_{\alpha3}(z) =  C_{\alpha3\beta3}(z) \varphi'_{\beta\gamma}(z)\gamma^{0^-}_{\gamma3} \label{eq:transverse_stresses} 
\end{equation}
For equilibrium reasons, these transverse stresses need to be continuous at the interfaces, and also to be null at $z = \pm h/2$ if the applied load remains normal to the plate. These conditions are expressed as 
\begin{subequations}
\begin{empheq}[left=\empheqlbrace]{align}
 &C_{\alpha3\gamma3}(\zeta_i^-)\varphi'_{\gamma\beta}(\zeta_i^-)=C_{\alpha3\gamma3}(\zeta_i^+)\varphi'_{\gamma\beta}(\zeta_i^+) \quad (i \in \{1,N-1\}) \\
 &C_{\alpha3\gamma3}(-h/2)\varphi'_{\gamma\beta}(-h/2)=0 \label{eq:bottom_stresses} \\
 &C_{\alpha3\gamma3}(+h/2)\varphi'_{\gamma\beta}(+h/2)=0 \label{eq:top_stresses}
\end{empheq}\label{eq:all_stresses}
\end{subequations}
We can pre-multiply equations~\eqref{eq:bottom_stresses} and~\eqref{eq:top_stresses} with the compliance tensors $S_{\delta3\alpha3}(-h/2)$ and $S_{\delta3\alpha3}(+h/2)$, respectively, to obtain: 
\begin{subequations}
\begin{empheq}[left=\empheqlbrace]{align}
     &\varphi'_{\delta\beta}(-h/2)=0\\
     &\varphi'_{\delta\beta}(+h/2)=0
\end{empheq}\label{eq:top_bottom_phi}
\end{subequations}
Figure~\ref{fig:Exemple3} illustrates the top and bottom nullity conditions and the jump conditions prescribed on the $\varphi'_{\alpha\beta}(z)$ functions. The following relations summarize all the properties that the functions $\varphi_{\alpha\beta}(z)$ are required to verify:
\begin{subequations}
\begin{empheq}[left=\empheqlbrace]{align}
     &\varphi_{\alpha\beta}(0)=0 \label{eq:prop1a} \\ 
     &\varphi_{\alpha\beta}(\zeta_i^-)=\varphi_{\alpha\beta}(\zeta_i^+), \quad (i \in \{1,N-1\}) \label{eq:prop1b} \\
     &\varphi'_{\alpha\beta}(0^-)=\delta_{\alpha\beta} \label{eq:prop1c} \\
     &C_{\alpha3\gamma3}(\zeta_i^-)\varphi'_{\gamma\beta}(\zeta_i^-)=C_{\alpha3\gamma3}(\zeta_i^+)\varphi'_{\gamma\beta}(\zeta_i^+)  \quad (i \in \{1,N-1\}) \label{eq:prop1d} \\
     &\varphi'_{\alpha\beta}(-h/2)=0 \label{eq:prop1e}\\   
     &\varphi'_{\alpha\beta}(+h/2)=0 \label{eq:prop1f}   
\end{empheq}\label{eq:properties_1}
\end{subequations}
According to the free indexes in the above formulas, we can see that there are $4+4(N-1)+4+4(N-1)+4+4=8(N+1)$ conditions for the four functions $\varphi_{\alpha\beta}$.
\section{The extension process\label{sec:ext}}

This Section describes the procedure for extending a generic bHSDT to a corresponding mHSDT. The construction of the $C_z^0$ \wfs{} is described and their tensorial character highlighted.
\subsection{Construction of the four $\varphi_{\alpha\beta}(z)$ functions}\label{sec:consfourwf}
Given a composite plate consisting of $N$ layers, the goal is to find four $\varphi_{\alpha\beta}(z)$ functions that obey to all the properties summarized in Eq.~\eqref{eq:properties_1}. Observing that the $\phi(z)$ function of the bHSDT is an odd function, one possibility would be to use it directly and to merely find an even $\gamma(z)$ function with suitable properties, in order to build the four \wfs{} from the basis spanned by the following $2N+2$ elements
\begin{equation}
(\;\phi(z),\,\gamma(z),\,\textbf{Z}_i(z),\,\textbf{1}_i(z)\;)
\end{equation}
$\textbf{Z}_i(z)$ and $\textbf{1}_i(z)$ represent the restrictions on the interval $]\zeta_{i-1},\zeta_i[$ of the linear and the constant (unitary) functions, respectively. Instead of $\phi(z)$ and $\gamma(z)$, the method proposed here uses more general and less constrained $f(z)$ and $g(z)$ functions, and hence is easier to use. The link that remains between the bHSDT and the corresponding mHSDT is the nature of the functions that will be used to form the basis (polynomial, trigonometric, hyperbolic\dots). In any case, these high-order functions are responsible for tailoring the transverse shear deformation, while the constant and linear elements introduce the characteristic zig-zag distribution of the in-plane displacements.
\par
We need two functions of class $C^1$: an odd function $f(z)$, and an even function $g(z)$ verifying $g'(\pm h/2) \ne 0$, {\em viz.}:
\begin{subequations}
\begin{empheq}[left=\empheqlbrace]{align}
     &f(z)=-f(-z) \label{eq:propfga} \\ 
     &g(z)= g(-z) \label{eq:propfgc} \\
     &g'(\pm h/2) \ne 0 \label{eq:propfge}
\end{empheq}\label{eq:properties_fg}
\end{subequations}
Now consider the four piecewise functions:
\begin{equation}
  \varphi_{\alpha\beta}(z)= a_{\alpha\beta}f(z)+b_{\alpha\beta}g(z)+c^{i}_{\alpha\beta}\textbf{Z}_i(z)+d^{i}_{\alpha\beta}\textbf{1}_i(z) \label{eq:WarpingFunctions}
\end{equation}
where summation is implied over the dummy index $i = 1,2,\ldots N$. These four functions are defined with $4(2+2N)=8(N+1)$ constants. The expression for the derivatives of the four functions is
\begin{equation}
  \varphi'_{\alpha\beta}(z)= a_{\alpha\beta}f'(z)+b_{\alpha\beta}g'(z)+c^{i}_{\alpha\beta}\textbf{1}_i(z)
  \label{eq:def_phi_prime}
\end{equation}
Just as the transverse shear strains, these four derivatives are not defined at the $N-1$ interfaces. %
Although Dirac's delta function might be used to formally write these derivatives, this is not useful because the relation in Eq.~\eqref{eq:prop1d} only involves their values at the layers' limits. %
Substituting Eqs.~\eqref{eq:WarpingFunctions} and \eqref{eq:def_phi_prime} into Eqs.~\eqref{eq:properties_1} yields the following system of $8(N+1)$ equations:
%
%
\begin{subequations}
\begin{empheq}[left=\empheqlbrace]{align}
     &d^{i_0}_{\alpha\beta}=0 \label{eq:prop2a} \\ 
     &c^{i}_{\alpha\beta}\zeta_i+d^{i}_{\alpha\beta}=c^{i+1}_{\alpha\beta}\zeta_i+d^{i+1}_{\alpha\beta} \quad (i \in \{1,N-1\}) \label{eq:prop2b} \\
     &a_{\alpha\beta}f'(0^-)+b_{\alpha\beta}g'(0^-)+c^{i_0}_{\alpha\beta}=\delta_{\alpha\beta} \label{eq:prop2c} \\ \nonumber
     &C^i_{\alpha3\gamma3}\left(a_{\gamma\beta}f'(\zeta_i)+b_{\gamma\beta}g'(\zeta_i)+c^{i}_{\gamma\beta}\right)\\
     &=C^{i+1}_{\alpha3\gamma3}\left(a_{\gamma\beta}f'(\zeta_i)+b_{\gamma\beta}g'(\zeta_i)+c^{i+1}_{\gamma\beta}\right)\quad (i \in \{1,N-1\}) \label{eq:prop2d} \\
     &a_{\alpha\beta}f'(-h/2)+b_{\alpha\beta}g'(-h/2)+c^{1}_{\alpha\beta}=0\label{eq:prop2e}\\
     &a_{\alpha\beta}f'(+h/2)+b_{\alpha\beta}g'(+h/2)+c^{N}_{\alpha\beta}=0\label{eq:prop2f}
\end{empheq}\label{eq:properties_2}
\end{subequations}
The index $i_0$ corresponds to the number of the layer which contains the $z=0^-$ coordinate. Since it seems difficult to formulate a recursive process to determine all the coefficients, the linear system~\eqref{eq:properties_2} is solved for the $8(N+1)$ unknown coefficients $a_{\alpha \beta}, b_{\alpha \beta}, c_{\alpha \beta}^i, d_{\alpha \beta}^i$ $(i=1,2\ldots N)$.
\par
Table~\ref{tab:Functions} reports some functions $f(z)$ and $g(z)$ that can be chosen to build an mHSDT model. While these functions allow to accommodate the transverse shear behaviour inside each layer, the linear and constant contributions are responsible for the ZZ behaviour, that is the respect of displacement and transverse stress continuities at the layers' interfaces. It should be noted that a ``mixed'' model can be constructed by using functions of different nature, for example the hyperbolic odd function $\sinh(kz/h)$ can be considered in conjunction with the even trigonometric function $\cos(\pi z/h)$. Analytical expressions for the \wfs{} for a single-layer plate are reported explicitly in Appendix.
\begin{table}[htbp]
	\centering\small
		\begin{tabular}{cccc}
	\hline \hline		
        Nature      &                $f(z)$                  &              $g(z)$                                                        &     Name
      \\ \hline 
       Polynomial   &                 $z^3$                  &               $z^2$                                                        &   \tozzfour{} 
      \\ \noalign{\smallskip}
      Trigonometric &   $\sin\left(\pi\dfrac{z}{h}\right)$   & $\cos\left(\pi\dfrac{z}{h}\right)$                                         &   \sizzfour{}
      \\ \noalign{\smallskip}
       Hyperbolic   &    $\sinh\left(k\dfrac{z}{h}\right)$   &   $\cosh\left(k\dfrac{z}{h}\right)$                                        &   \hyzzfour{}
      \\ \noalign{\smallskip}
	\hline \hline		
		\end{tabular}
	\caption{Nature of the original bHSDT and corresponding couple of functions $f(z)$ and $g(z)$ used to build the mHSDT.}
	\label{tab:Functions}
\end{table}
%
%
%
\subsection{Computational aspects}
The construction of the linear system \eqref{eq:properties_2} can be automated because its structure does not depend on the choice of the functions $f(z)$ and $g(z)$. Indeed, only few values of these functions and of their derivatives, taken at specific $z$ coordinates, have to be sent to the routine. The solution of the $(8N+8)\times(8N+8)$ linear algebraic system can be carried out with a classical algorithm and provides the coefficients defining the four \wfs{} $\varphi_{\alpha\beta}(z)$. It may be noted that either numerical or semi-analytical versions of the \wfs{} can be used. Numerical versions, which consist on a sufficiently dense table of values, are more suitable for computing the numerous generalized stiffness and mass terms of the plate model within a  numerical quadrature scheme.
\subsection{Stress functions}
Once the $\varphi_{\alpha\beta}(z)$ functions are built, one can compute the corresponding \sfs{} $\psi_{\alpha\beta}(z)$. They do not bring new information to the models, as these \sfs{} are a direct consequence of the \wfs{}, but they are useful to illustrate and understand the static response of the mHSDT. Let us replace in equation~\eqref{eq:transverse_stresses}, the middle-plane transverse strains by the corresponding middle-plane stresses:
\begin{equation}
  \sigma_{\alpha3}(z) =  C_{\alpha3\beta3}(z)\, \varphi'_{\beta\gamma}(z)\,\gamma^{0^-}_{\gamma3} = 4\,C_{\alpha3\beta3}(z) \,\varphi'_{\beta\gamma}(z)\,S_{\gamma3\delta3}(0^-) \, \sigma^{0}_{\delta3} \label{eq:stress_functions}
\end{equation}
Note that the $0^-$ exponent of $\gamma^{0^-}_{\gamma3}$ is not required to appear in the interlaminar continuous stress $\sigma^{0}_{\delta3}$, but it is found in the $S_{\gamma3\delta3}(0^-)$ term.
Eq.~\eqref{eq:stress_functions} permits to define the 4 \sfs{} of the model
\begin{equation}
  \psi_{\alpha\beta}(z) =  4 \,C_{\alpha3\gamma3}(z) \,\varphi'_{\gamma\delta}(z) \,S_{\delta3\beta3}(0^-) \label{eq:stress_functions2}
\end{equation}%
%
through which the transverse shear stresses are expressed as
\begin{equation}
  \sigma_{\alpha3}(z) = \psi_{\alpha\beta}(z)\,\sigma^{0}_{\beta3}  \label{eq:stress_functions3}
\end{equation}
%
%
\subsection{Tensorial character of the $\varphi_{\alpha\beta}(z)$}
The tensorial character of the $\varphi_{\alpha\beta}(z)$ functions follows directly from their definition, see Eq.~\eqref{eq:mHSDT}. This tensorial character concerns only the 2D $(x,y)$ space. Also, the equations of the system~\eqref{eq:properties_1} are tensor equations, i.e., their form is invariant with respect to rotations about the $z$ axis. It implies that all the coefficients $a_{\alpha\beta}$, $b_{\alpha\beta}$, $c^i_{\alpha\beta}$, $d^i_{\alpha\beta}$ are second order tensors and must, therefore, obey to the formulas of coordinate transformation for second order tensors. 
%
\par
The tensorial character of the \wfs{} implies that the four functions $\varphi_{\alpha\beta}(z)$ of a laminate whose lamination sequence is $s=[\theta_1/\theta_2/\dots/\theta_N]$ must be linked to the four functions $\bar{\varphi}_{\alpha\beta}(z)$ of the laminate whose stacking sequence is $\bar{s}=s+\theta=[(\theta_1+\theta)/(\theta_2+\theta)/\dots/(\theta_N+\theta)]$. 

In order to identify this relation, let us consider the $\bar{s}$--laminate, in a Cartesian frame $(x,y,z)$, and suppose it undergoes a pure shear deformation of its middle plane $\bar{\pmb{\upgamma}}$. In this case, the kinematic field of Eq.~\eqref{eq:in_plane_depl} can be written $\bar{u}_{\alpha}(z)=\bar{\varphi}_{\alpha\beta}(z) \bar{\gamma}^0_{\beta3}$ or, in matrix notation, $\bar{\textbf{u}}=\bar{\pmb{\upvarphi}} \bar{\pmb{\upgamma}}$. Consider the matrix of change of coordinates $\textbf{P}_\theta$ from the Cartesian frame $(x,y,z)$ to a Cartesian frame $(x',y',z)$, rotated from the previous one by an angle $\theta$ about the $z$-axis. In the rotated frame, this shear strain is $\pmb{\upgamma}=\textbf{P}_\theta\bar{\pmb{\upgamma}}$ and it ``acts'' on the $s$--laminate producing the in-plane kinematic field $\textbf{u}$. In the original frame, the $\bar{s}$--laminate is then subjected to the kinematic field $\bar{\textbf{u}}=\textbf{P}_\theta^{-1}\textbf{u}$. Therefore, the following relation is established: $\bar{\pmb{\upvarphi}}=\textbf{P}_\theta^{-1}\pmb{\upvarphi}\textbf{P}_\theta$. 
%
%
Since the transformation matrix is
\begin{equation}
  \textbf{P}_\theta=
  \left[
  \begin{array}{cc} 
     c & s  \\
    -s & c
  \end{array}
  \right] 
  \quad \text{with} ~ c=\cos(\theta) ~ \text{and} ~ s=\sin(\theta)  
\end{equation}
one can compute the \wfs{} for the $\bar{s}$--laminate directly from those for the $s$--laminate according to
\begin{equation}\label{eq:tensTrans}
  \left\{
  \begin{array}{l} 
    \bar{\varphi}_{11}=  \varphi_{11}c^2 + \varphi_{22}s^2 + (\varphi_{12}    + \varphi_{21})sc \\ 
    \bar{\varphi}_{12}=-(\varphi_{11}    - \varphi_{22})sc +  \varphi_{12}c^2 - \varphi_{21}s^2 \\ 
    \bar{\varphi}_{21}=-(\varphi_{11}    - \varphi_{22})sc -  \varphi_{12}s^2 + \varphi_{21}c^2 \\ 
    \bar{\varphi}_{22}=  \varphi_{11}s^2 + \varphi_{22}c^2 - (\varphi_{12}    + \varphi_{21})sc
  \end{array}
  \right. 
\end{equation}
\par
Two examples of such transformations are given next for illustration purposes. Figure~\ref{fig:Tensorial_p0p90_phi} compares the native \wfs{} of a $[45/{-}45]$ laminate against those obtained from a $[0/90]$ laminate after rotating them by an angle of $45^{\circ}$. The same comparison is proposed in figure~\ref{fig:Tensorial_m10p0p40_phi} for the two laminates $[-25/{-}15/25]$ and $[-10/0/40]$ and with a rotation of $-15^{\circ}$.
\iftoggle{submission}{}{\tikzsetnextfilename{Tensorial_p0p90_phi}}
\begin{SmartFigure}[htbp]
  \centering
  \iftoggle{submission}{
    \includegraphics{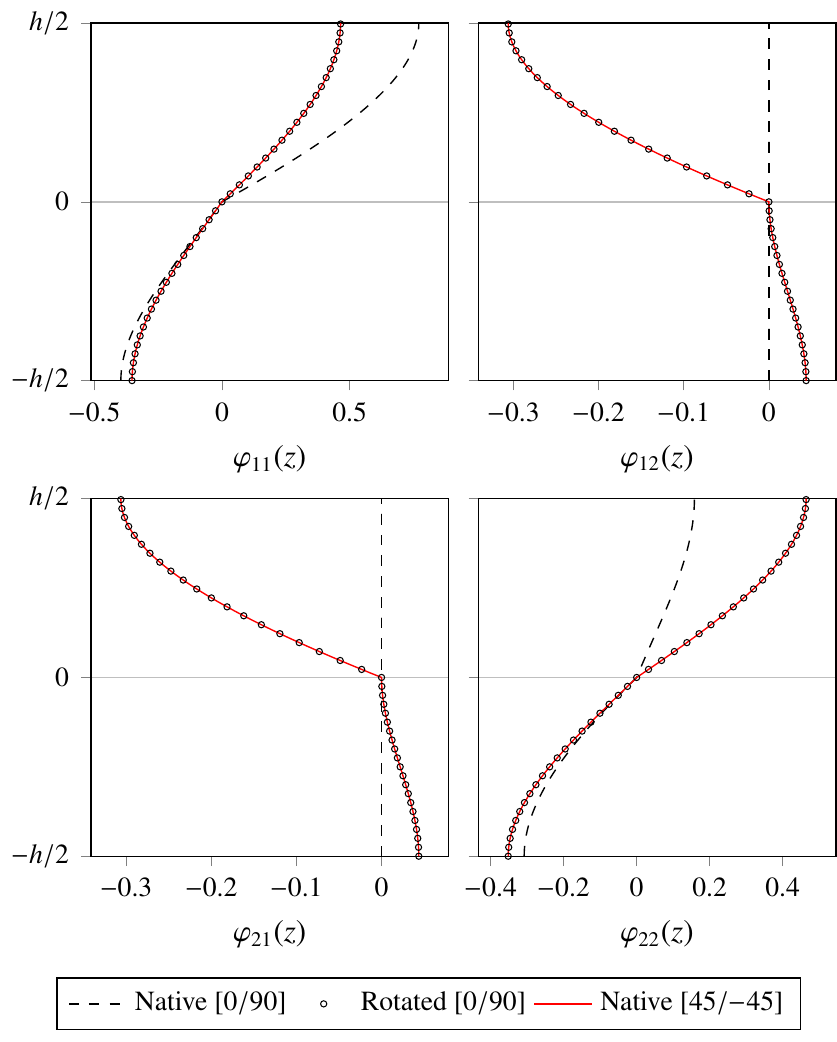}
  }{
    \input{Figures/Tensorial_p0p90_phi.tex}
  }
  \caption{Tensorial behaviour of the \wfs{}: comparison between the native \wfs{} of a $[45/{-45}]$ laminate and those obtained by the coordinate transformation from a $[0/90]$ laminate.}%
  \label{fig:Tensorial_p0p90_phi}
\end{SmartFigure}
\iftoggle{submission}{}{\tikzsetnextfilename{Tensorial_m10p0p40_phi}}
\begin{SmartFigure}[htbp]
  \centering
  \iftoggle{submission}{
    \includegraphics{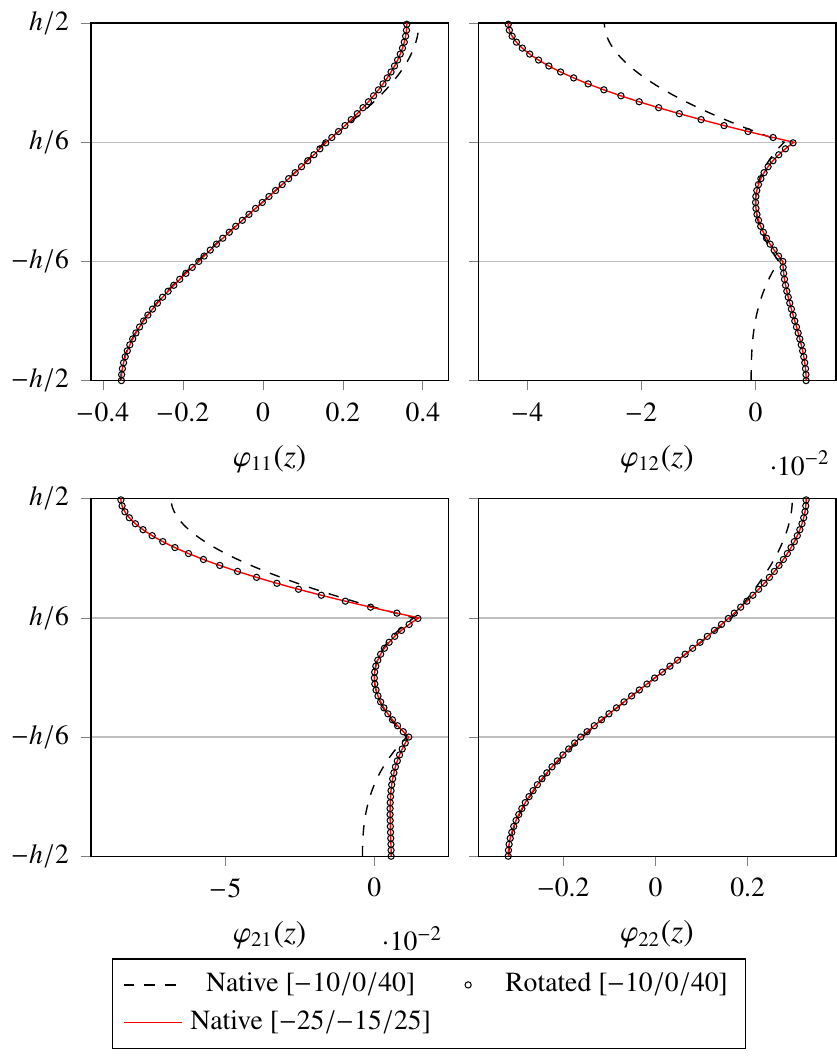}
  }{
    \input{Figures/Tensorial_m10p0p40_phi.tex}
  }
  \caption{Tensorial behaviour of the \wfs{}: comparison between the native polynomial $[-25/{-15}/25]$ functions and those obtained by the coordinate transformation of the $[-10/0/40]$.}%
  \label{fig:Tensorial_m10p0p40_phi}
\end{SmartFigure}
\section{Numerical results}\label{sec:results}
A numerical evaluation is proposed in order to assess the accuracy of the basis models and their corresponding enhancement through ZZ \wfs{} with respect to the plate's length-to-thickness ratio, number of layers, and stacking sequence. All bHSDT listed in Table~\ref{tab:bHSDTFunctions} are compared with their corresponding enhancements defined by the functions listed in Table~\ref{tab:Functions}. The factor $k$ in the hyperbolic functions has been set equal to 2 in the subsequent numerical investigations. Note finally that all considered models have the same number of DOF as the bHSDT, i.e., 5 DOF. 

In order to encompass a quite broad range of stiffness mismatch between adjacent layers, the study will investigate laminated as well as sandwich plates with composite skins and a honeycomb core. The material properties used for the composite and the honeycomb layers are reported in Table~\ref{tab:matprop}.

The numerical assessment of the different models is carried out by referring to an exact solution of the 2D differential equations governing the plate bending problem. Square, simply-supported plates are considered, for which we compute the fundamental eigenfrequency as well as the static response under bi-sinusoidal transverse pressure loads of amplitude $q/2$ acting at the top and the bottom surfaces of the plate. A Navier-type strong-form solution is found for all considered laminates, where the simply-support conditions of arbitrary, non cross-ply laminates are opportunely relaxed as discussed in \cite{Loredo-3D-COST2014}. The \wfs{} and \sfs{} of the bHSDT and mHSDT are compared against those that have been extracted from the 3D solutions following the procedure detailed out in \cite{Loredo2014}.

The points at which quantities are output are defined as follows:
\begin{equation}\begin{split}
A=(a/2,0,0); &\quad A'=(a/2,0,h/4)\\
B=(0,a/2,0); &\quad B'=(0,a/2,h/4)\\
C=(a/2,a/2,0); &\quad C^+=(a/2,a/2,h/2)
\end{split}\end{equation} 
Deflections $w$, first natural frequencies $\omega$ and stresses are given according to following adimensionalisation
  \begin{equation}\begin{split}
    \overline{w}=100\frac{E_2^{\text{ref}} h^3}{q\,a^4}\,w 
    \,\text{,}&\quad
    \overline{\omega}=\frac{a^2}{h}\sqrt{\frac{\rho^{\text{ref}}}{E_2^{\text{ref}}}}\,\omega
\\
    \overline{\sigma}_{\alpha \beta}= 10 \frac{h^2}{q\,a^2} \, \sigma_{\alpha \beta}
    \,\text{,}&\quad
    \overline{\sigma}_{\alpha 3}=10\frac{h}{q\,a} \, \sigma_{\alpha 3}
	\label{eq:nondimd}
\end{split}\end{equation}
where for sandwich plates $E_2^{\text{ref}}$ and $\rho^{\text{ref}}$ are the values of the core material. It is important to specify that the transverse shear stress values reported in the tables and their distributions across the plate thickness plotted in the figures are obtained from the equilibrium equations upon integrating the in-plane stresses.
\begin{table}[htbp]
  \centering
  \footnotesize
  \setlength{\tabcolsep}{1.5pt}
  \begin{tabular}{lccccccccc}
  \hline \hline
                       &   $E_1$   &    $E_2$   &    $E_3$    &  $G_{23}$  &  $G_{13}$  &  $G_{12}$  & $\nu_{\alpha 3}$ & $\nu_{12}$ &  $\rho$  \\[4pt]	
   \hline
    Composite (c)      & $25E^c_2$ &   $E^c_2$  &   $E^c_2$   & $0.2E^c_2$ & $0.5E^c_2$ & $0.5E^c_2$ &   $0.25$   &   $0.25$   & $\rho^c$ \\[4pt]
    Honeycomb (h)      &  $E^h_2$  & $E^c_2/25$ & $12.5E^h_2$ & $1.5E^h_2$ & $1.5E^h_2$ & $0.4E^h_2$ &   $0.02$   &   $0.25$   &  $\rho^c/15$ \\[4pt]
  \hline \hline
	\end{tabular}
	\caption{Material properties ($\alpha= 1,2$)}
	 \label{tab:matprop}
\end{table}
\subsection{The $[0/90]_n$ laminates}
The models are tested for the antisymmetric cross-ply laminates $[0/90]_n$, where different numbers of layers are considered with $n=1,2,3,4,5,10$. In table~\ref{tab:p0p90n}, non-dimensional deflection, transverse shear stresses and fundamental frequency are given for the Sin and the \sizzfour{} models, and compared to the exact solution. The length-to-thickness ratio is set to $a/h=10$. Very similar results are obtained with polynomial and hyperbolic bHSDT/mHSDT models and are omitted from Table~\ref{tab:p0p90n} for the sake of clarity. The results clearly shows the accuracy improvement introduced by the $C^0_z$ \wfs{}, in particular for the deflection and the fundamental frequency: the enhancement on these two quantities appears to decrease as the number of layers increases, although for $n=10$ it is still larger than 5\% and 3\%, respectively. 

Only two \wfs{} are required for a cross-ply laminate because the cross-coupling functions are identically nil, $\varphi_{12}(z) = \varphi_{21}(z) = 0$. The functions $\varphi_{11}(z)$ and $\varphi_{22}(z)$ of the polynomial bHSDT (ToSDT) and of the polynomial, trigonometric and hyperbolic mHSDT are compared in Figures~\ref{fig:p0p90x2_phi} and \ref{fig:p0p90x4_phi} for the $n=2$ and $n=4$ configurations, respectively. The differences between the 3 mHSDT are seen to be negligible, and the curves for the trigonometric and hyperbolic bHSDT have been omitted for the sake of clarity because they are practically coincident with those of the ToSDT.

As far as the impact of \wfs{} on the local stress response is concerned, the values in Table~\ref{tab:p0p90n} for the transverse shear stresses at the selected points do not allow to well appreciate it, but their through-the-thickness distributions obtained with the extended mHSDT model are closer to the exact solution in comparison to the bHSDT models. This can be seen by comparing the two \sfs{} $\psi_{11}(z)$ and $\psi_{22}(z)$ depicted in Figures~\ref{fig:p0p90x2_psi} and \ref{fig:p0p90x4_psi} for the cases $n=2$ and $n=4$, respectively. On the other hand, Figures~\ref{fig:p0p90x2_sig} and \ref{fig:p0p90x4_sig} report the transverse shear stress distributions computed for the two configurations $n=2$ and $n=4$, respectively, upon integrating the equilibrium equations starting from the in-plane stresses. This post-processing procedure is seen to annihilate all differences between the bHSDT and the mHSDT, thus providing distributions that very accurately recover the exact 3D solution.

All considered models for the cross-ply laminate $[0/90]_2$ are assessed in Table~\ref{tab:p0p90x2} with respect to the length-to-thickness ratio $a/h$. It is obvious that the improvement of the mHSDT over the bHSDT is more important for thick plates than for thin plates, it decreases from more than 10\% for $a/h=4$ to about 0.1\% for $a/h=100$. This is not surprising as it is well known that the effect of the transverse shear increases as $a/h$ diminishes.

\par

It is worthwhile to make some comments about an expected symmetry for the \wfs{} of the considered antisymmetric cross-ply laminates. Indeed, one should expect the $\varphi_{11}(z)$ functions to be equal to the corresponding $\varphi_{22}(-z)$ functions, but figures~\ref{fig:p0p90x2_phi} and~\ref{fig:p0p90x4_phi} show that this is not the case. This is due to the fact that the $z=0^-$ coordinate has been chosen for prescribing the $\varphi'_{11}(0^-)=\varphi'_{22}(0^-)=1$ conditions. This choice ``hides'' the expected property, which can nevertheless be easily restored: dividing $\varphi_{22}(-z)$ by $\varphi'_{22}(0^+)$ yields in fact exactly the function $\varphi_{11}(z)$. Note that, since there are no such constraints on the \sfs{}, the symmetry $\psi_{11}(z)=\psi_{22}(-z)$ is immediately apparent in Figures~\ref{fig:p0p90x2_psi} and~\ref{fig:p0p90x4_psi}.

\iftoggle{submission}{}{\tikzsetnextfilename{p0p90x2_phi}}
\begin{SmartFigure}[htbp]
  \centering
  \iftoggle{submission}{
    \includegraphics{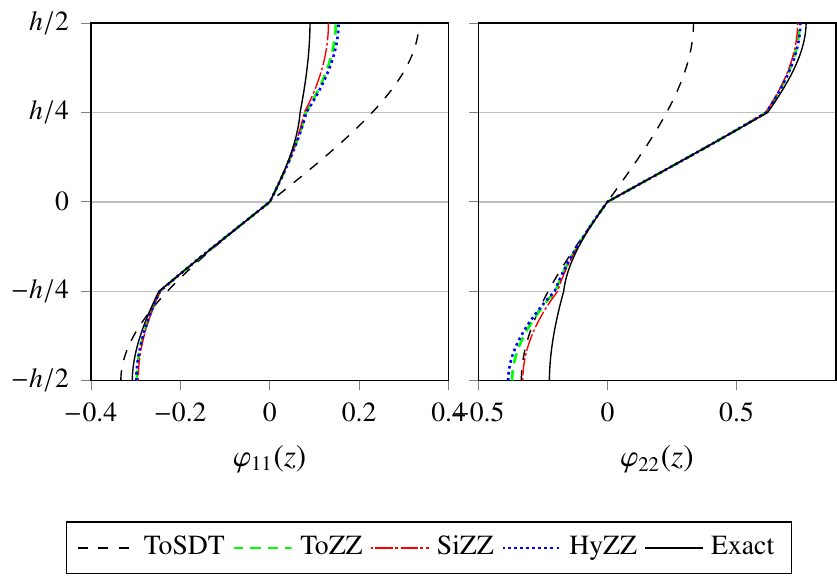}
  }{
    \input{Figures/p0p90x2_phi.tex}
  }
  \caption{\Wfs{} of the $[0/90]_2$ square plate with $a/h=10$ for each considered model.}%
  \label{fig:p0p90x2_phi}
\end{SmartFigure}
\iftoggle{submission}{}{\tikzsetnextfilename{p0p90x2_stress_func}}
\begin{SmartFigure}[htbp]
  \centering
  \iftoggle{submission}{
    \includegraphics{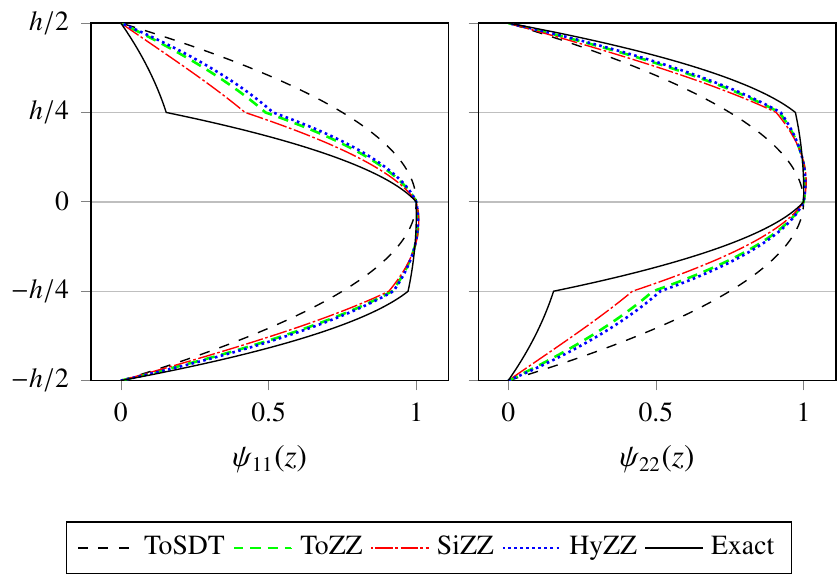}
  }{
    \input{Figures/p0p90x2_stress_func.tex}
  }
  \caption{Transverse shear \sfs{} of the $[0/90]_2$ square plate with $a/h=10$ for each considered model.}%
  \label{fig:p0p90x2_psi}
\end{SmartFigure}
\iftoggle{submission}{}{\tikzsetnextfilename{p0p90x2_sig}}
\begin{SmartFigure}[htbp]
  \centering
  \iftoggle{submission}{
    \includegraphics{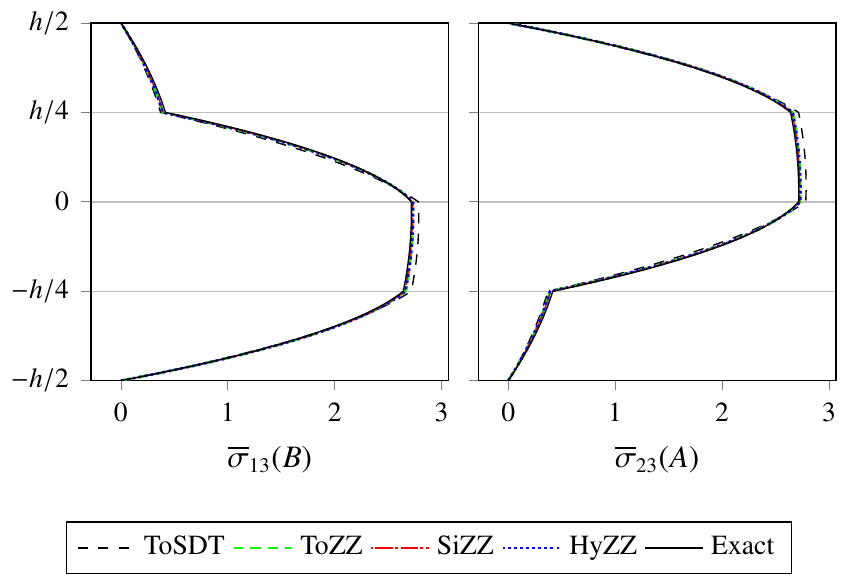}
  }{
    \input{Figures/p0p90x2_sig.tex}
  }
  \caption{Post-processed transverse shear stresses of the $[0/90]_2$ square plate with $a/h=10$ for each considered model.}%
  \label{fig:p0p90x2_sig}
\end{SmartFigure}
\iftoggle{submission}{}{\tikzsetnextfilename{p0p90x4_phi}}
\begin{SmartFigure}[htbp]
  \centering
  \iftoggle{submission}{
    \includegraphics{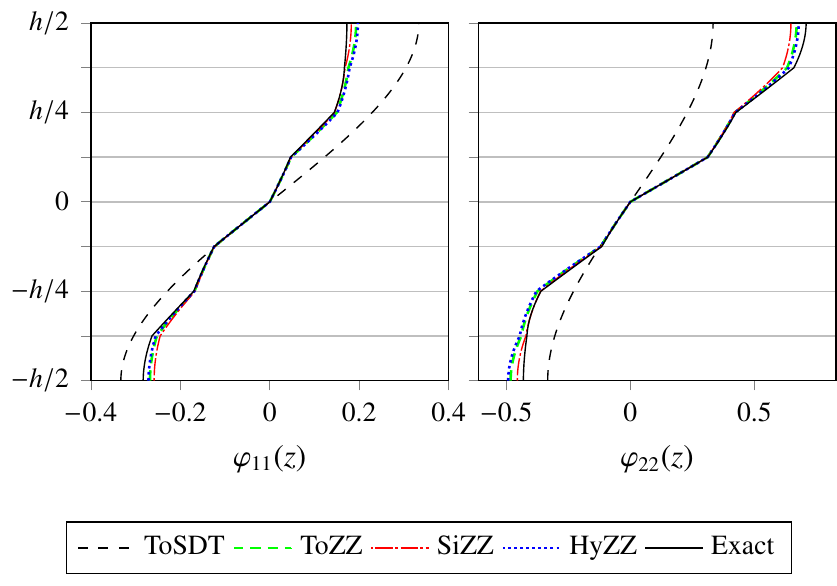}
  }{
    \input{Figures/p0p90x4_phi.tex}
  }
  \caption{\Wfs{} of the $[0/90]_4$ square plate with $a/h=10$ for each considered model.}%
  \label{fig:p0p90x4_phi}
\end{SmartFigure}
\iftoggle{submission}{}{\tikzsetnextfilename{p0p90x4_stress_func}}
\begin{SmartFigure}[htbp]
  \centering
  \iftoggle{submission}{
    \includegraphics{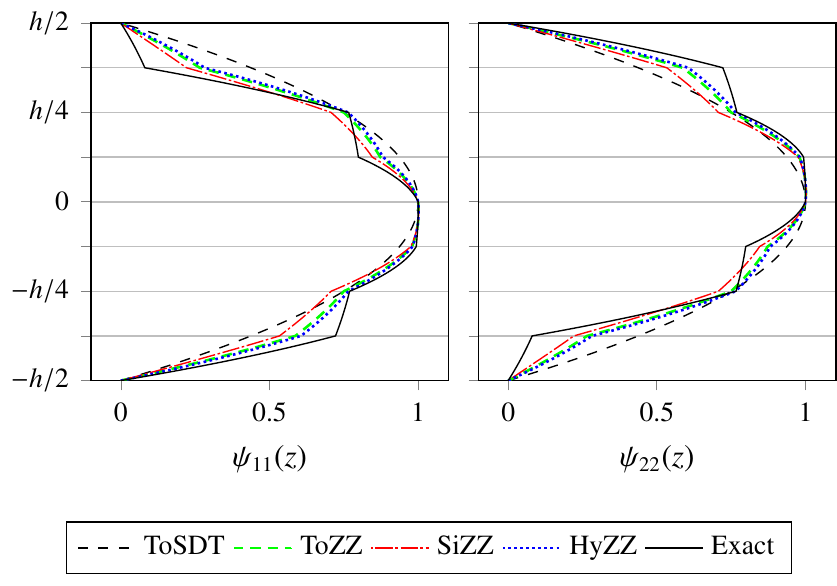}
  }{
    \input{Figures/p0p90x4_stress_func.tex}
  }
  \caption{Transverse shear \sfs{} of the $[0/90]_4$ square plate with $a/h=10$ for each considered model.}%
  \label{fig:p0p90x4_psi}
\end{SmartFigure}
\iftoggle{submission}{}{\tikzsetnextfilename{p0p90x4_sig}}
\begin{SmartFigure}[htbp]
  \centering
  \iftoggle{submission}{
    \includegraphics{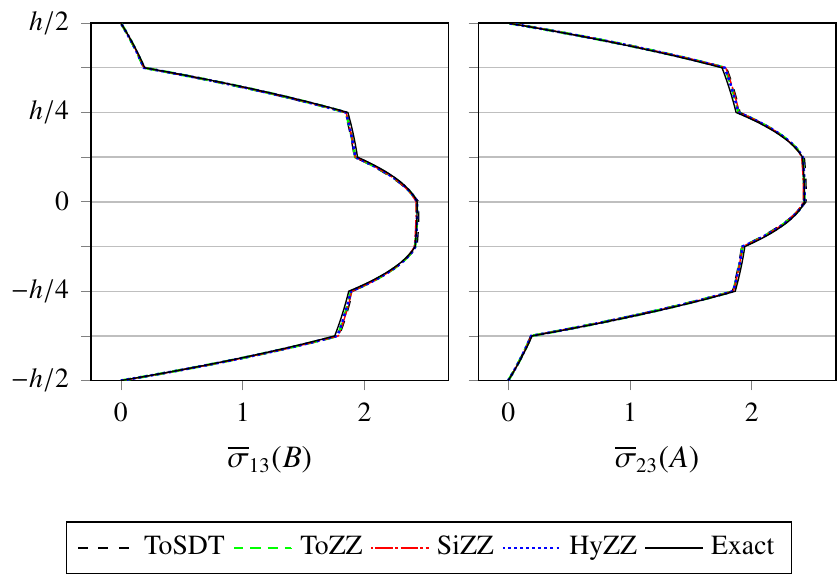}
  }{
    \input{Figures/p0p90x4_sig.tex}
  }
  \caption{Post-processed transverse shear stresses of the $[0/90]_4$ square plate with $a/h=10$ for each considered model.}%
  \label{fig:p0p90x4_sig}
\end{SmartFigure}
\begin{table}[htbp]%
	\centering
	\footnotesize
  \setlength{\tabcolsep}{3pt}
	\begin{tabular}{c@{~}l|c@{~}c|l@{~}c|c@{~}c|c@{~}c}
    \iftoggle{submission}{
      Seq. & Model & $\overline{w}(C)$ & \% & $\overline{\sigma}_{13}(B)$ & \% & $\overline{\sigma}_{22}(C^+)$ & \% & $\overline{\omega}$ & \% \\
      \hline
      $[0/90]$ & Sin         & $1.2132$ & $-1.17$ & $1.1990$ & $-2.84$ & $7.4827$ & $+2.41$ & $8.9869$ & $+0.72$ \\ 
         & \sizzfour{} & $1.2046$ & $-1.87$ & $1.1792$ & $-4.45$ & $7.4983$ & $+2.62$ & $9.0204$ & $+1.10$ \\ 
         & Exact       & $1.2275$ &  & $1.2341$ &  & $7.3065$ &  & $8.9226$ &  \\
      \hline
      $[0/90]_2$ & Sin         & $0.68725$ & $-9.86$ & $2.7806$ & $+2.37$ & $5.2317$ & $-1.43$ & $11.998$ & $+5.41$ \\ 
           & \sizzfour{} & $0.75863$ & $-0.49$ & $2.7250$ & $+0.33$ & $5.3714$ & $+1.21$ & $11.425$ & $+0.38$ \\ 
           & Exact & $0.76239$ &  & $2.7161$ &  & $5.3074$ &  & $11.382$ &  \\
      \hline
      $[0/90]_3$ & Sin         & $0.63818$ & $-7.91$ & $2.2818$ & $-0.87$ & $5.2357$ & $+0.64$ & $12.459$ & $+4.28$ \\ 
           & \sizzfour{} & $0.69191$ & $-0.16$ & $2.2781$ & $-1.03$ & $5.2907$ & $+1.70$ & $11.972$ & $+0.20$ \\ 
           & Exact       & $0.69303$ &  & $2.3018$ &  & $5.2026$ &  & $11.948$ &  \\
      \hline
      $[0/90]_4$ & Sin         & $0.62262$ & $-7.05$ & $2.4403$ & $+0.36$ & $5.3053$ & $+1.12$ & $12.617$ & $+3.79$ \\ 
           & \sizzfour{} & $0.66969$ & $-0.02$ & $2.4229$ & $-0.36$ & $5.3433$ & $+1.84$ & $12.171$ & $+0.12$ \\ 
           & Exact       & $0.66982$ &  & $2.4316$ &  & $5.2467$ &  & $12.156$ &  \\
      \hline
      $[0/90]_5$ & Sin         & $0.61566$ & $-6.61$ & $2.3214$ & $-0.35$ & $5.3673$ & $+1.17$ & $12.689$ & $+3.54$ \\ 
           & \sizzfour{} & $0.65965$ & $+0.06$ & $2.3138$ & $-0.67$ & $5.4030$ & $+1.84$ & $12.265$ & $+0.08$ \\ 
           & Exact       & $0.65923$ &  & $2.3294$ &  & $5.3053$ &  & $12.255$ &  \\
      \hline
      $[0/90]_{10}$ & Sin         & $0.60663$ & $-5.99$ & $2.3575$ & $+0.06$ & $5.5367$ & $+0.71$ & $12.785$ & $+3.20$ \\ 
              & \sizzfour{} & $0.64650$ & $+0.19$ & $2.3464$ & $-0.41$ & $5.5853$ & $+1.59$ & $12.390$ & $+0.02$ \\ 
              & Exact       & $0.64530$ &  & $2.3561$ &  & $5.4979$ &  & $12.388$ &  \\
   }{
      
      \hline
      
      \hline
      
      \hline
      
      \hline
      
      \hline
      
      \hline
      
   }  
	\end{tabular}
	\normalsize
	\caption{Comparison between the Sin and the \protect\sizzfour{} models for the $[0/90]_n$ with $n=1,2,3,4,5,10$ ($a/h=10$).}
	\label{tab:p0p90n}
\end{table}
\begin{table}[htbp]%
	\centering
	\footnotesize
  \setlength{\tabcolsep}{3pt}
	\begin{tabular}{c@{~}l|c@{~}c|l@{~}c|c@{~}c|c@{~}c}
    \iftoggle{submission}{
      a/h & Model & $\overline{w}(C)$ & \% & $\overline{\sigma}_{13}(B)$ & \% & $\overline{\sigma}_{22}(C^+)$ & \% & $\overline{\omega}$ & \% \\ 

      \hline
      4 & ToSDT       & $1.6093$ & $-17.8$ & $2.6745$ & $+12.9$ & $6.9699$ & $-3.63$ & $7.8250$ & $+11.4$ \\ 
  & \tozzfour{} & $1.9484$ & $-0.49$ & $2.3752$ & $+0.30$ & $7.7158$ & $+6.69$ & $7.1053$ & $+1.13$ \\ 
  & Sin         & $1.6091$ & $-17.8$ & $2.6605$ & $+12.3$ & $7.0862$ & $-2.02$ & $7.8237$ & $+11.3$ \\ 
  & \sizzfour{} & $1.9571$ & $-0.05$ & $2.3490$ & $-0.81$ & $7.7644$ & $+7.36$ & $7.0847$ & $+0.84$ \\ 
  & Hyp         & $1.6086$ & $-17.8$ & $2.6794$ & $+13.1$ & $6.9271$ & $-4.22$ & $7.8273$ & $+11.4$ \\ 
  & \hyzzfour{} & $1.9434$ & $-0.75$ & $2.3852$ & $+0.72$ & $7.6943$ & $+6.39$ & $7.1161$ & $+1.28$ \\ 
  & Exact & $1.9581$ &  & $2.3681$ &  & $7.2321$ &  & $7.026$ &  \\ 

      \hline
      10 & ToSDT       & $0.68655$ & $-9.95$ & $2.7829$ & $+2.46$ & $5.2112$ & $-1.81$ & $12.004$ & $+5.47$ \\ 
   & \tozzfour{} & $0.75496$ & $-0.97$ & $2.7302$ & $+0.52$ & $5.3587$ & $+0.97$ & $11.453$ & $+0.63$ \\ 
   & Sin         & $0.68725$ & $-9.86$ & $2.7806$ & $+2.37$ & $5.2317$ & $-1.43$ & $11.998$ & $+5.41$ \\ 
   & \sizzfour{} & $0.75863$ & $-0.49$ & $2.7250$ & $+0.33$ & $5.3714$ & $+1.21$ & $11.425$ & $+0.38$ \\ 
   & Hyp         & $0.68619$ & $-9.99$ & $2.7838$ & $+2.49$ & $5.2038$ & $-1.95$ & $12.007$ & $+5.49$ \\ 
   & \hyzzfour{} & $0.75336$ & $-1.18$ & $2.7321$ & $+0.59$ & $5.3536$ & $+0.87$ & $11.466$ & $+0.73$ \\ 
   & Exact & $0.76239$ &  & $2.7161$ &  & $5.3074$ &  & $11.382$ &  \\ 

      \hline
      100 & ToSDT       & $0.50834$ & $-0.16$ & $2.8039$ & $+0.03$ & $4.8715$ & $-0.02$ & $14.024$ & $+0.08$ \\ 
    & \tozzfour{} & $0.50906$ & $-0.02$ & $2.8034$ & $+0.01$ & $4.8731$ & $+0.01$ & $14.014$ & $+0.01$ \\ 
    & Sin         & $0.50835$ & $-0.16$ & $2.8039$ & $+0.02$ & $4.8718$ & $-0.02$ & $14.024$ & $+0.08$ \\ 
    & \sizzfour{} & $0.50910$ & $-0.01$ & $2.8033$ & $+0.00$ & $4.8732$ & $+0.01$ & $14.014$ & $+0.01$ \\ 
    & Hyp         & $0.50834$ & $-0.16$ & $2.8039$ & $+0.03$ & $4.8715$ & $-0.02$ & $14.024$ & $+0.08$ \\ 
    & \hyzzfour{} & $0.50904$ & $-0.02$ & $2.8034$ & $+0.01$ & $4.8730$ & $+0.01$ & $14.015$ & $+0.01$ \\ 
    & Exact & $0.50915$ &  & $2.8032$ &  & $4.8727$ &  & $14.013$ &  \\ 

   }{
      
      \hline
      
      \hline
      
      \hline
      
   }  
	\end{tabular}
	\normalsize
	\caption{Comparison between the different models for the square $[0/90]_2$ plate with a varying length to thickness ratio.}
	\label{tab:p0p90x2}
\end{table}

\subsection{The $[30/{-30}]_n$ laminates}
The $[30/{-30}]_n$ laminate family is next considered, for which all four \wfs{} are required due to the off-axis orientation angles. For the moderately thick plate characterized by $a/h=10$ with $n=4$, figure~\ref{fig:p30m30x4_phi} compares the \wfs{} $\varphi_{\alpha\beta}(z)$ of the \tosdt{} and of the 3 mHSDT with those obtained from the 3D solution; the corresponding \sfs{} $\psi_{\alpha\beta}(z)$ are plotted in figure~\ref{fig:p30m30x4_psi}. The differences between the three mHSDT are again negligible. Even if the discrepancy with respect to the exact \sfs{} may be relevant, very accurate transverse shear stresses are obtained from the integration of the equilibrium equations, as shown in figure~\ref{fig:p30m30x4_sig}. 
\par
The numerical values for deflection, stresses and fundamental frequency are reported in tables~\ref{tab:p30m30n} and \ref{tab:p30m30x2}. The results in table~\ref{tab:p30m30n} refer to the trigonometric models Sin/\sizzfour{} for a fixed length-to-thickness ratio $a/h=10$ and for different numbers of layers $n=1,2,3,4,5,10$. Note that the relative errors for $\sigma_{23}(B')$ and $\sigma_{13}(A')$ for $n=4$ have not been reported because the values of the exact solution are very small. Table~\ref{tab:p30m30x2} compares the considered mHSDT for various ratios $a/h$ and the fixed $n=2$ case. The same conclusions can be drawn as those concerning the cross-ply $[0/90]_n$ laminates: exception made for the 2-layer $n=1$ case, mHSDT models substantially improve the corresponding bHSDT models, and this difference is more significant if the $a/h$ ratio is low. The improvement is more systematic for the global response than for the local stress response, for which a certain dependency is observed with respect to the number $n$ of the stacking sequence.
\iftoggle{submission}{}{\tikzsetnextfilename{p30m30x4_phi}}
\begin{SmartFigure}[htbp]
  \centering
  \iftoggle{submission}{
    \includegraphics{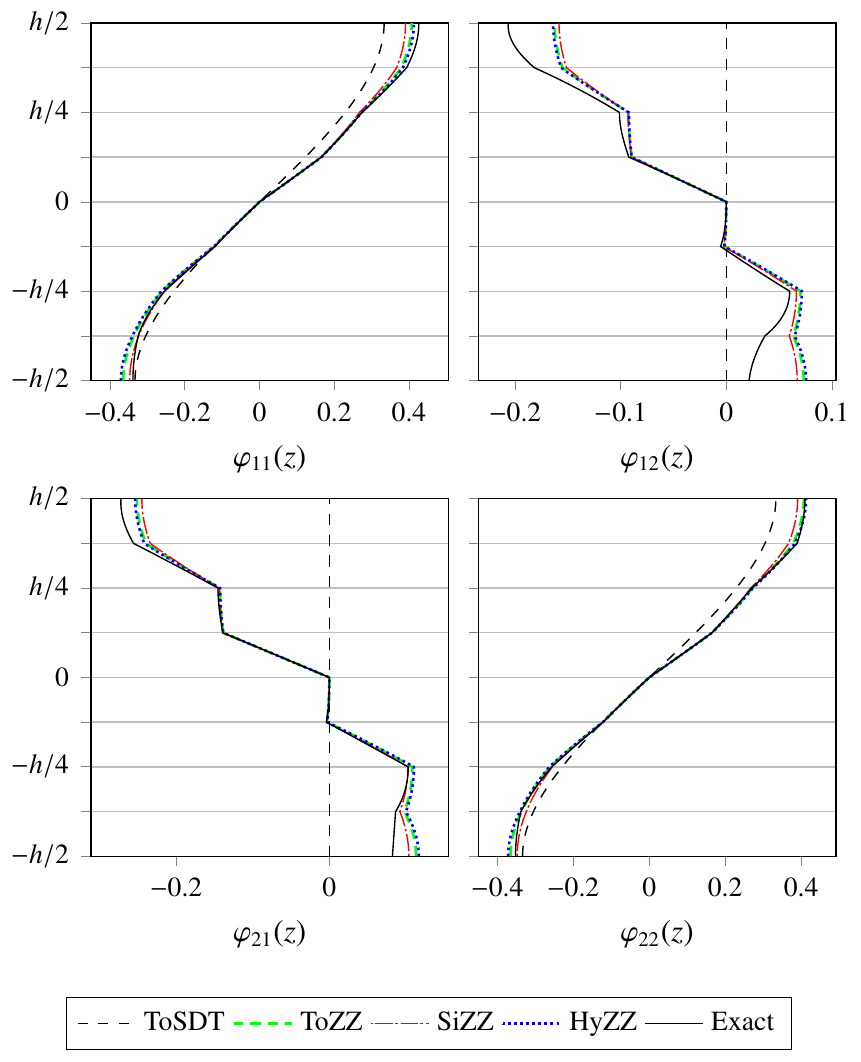}
  }{
    \input{Figures/p30m30x4_phi.tex}%
  }
  \caption{The four \wfs{} for the $[30/{-30}]_4$ laminate with $a/h=10$}%
  \label{fig:p30m30x4_phi}
\end{SmartFigure}
\iftoggle{submission}{}{\tikzsetnextfilename{p30m30x4_stress_func}}
\begin{SmartFigure}[htbp]
  \centering
  \iftoggle{submission}{
    \includegraphics{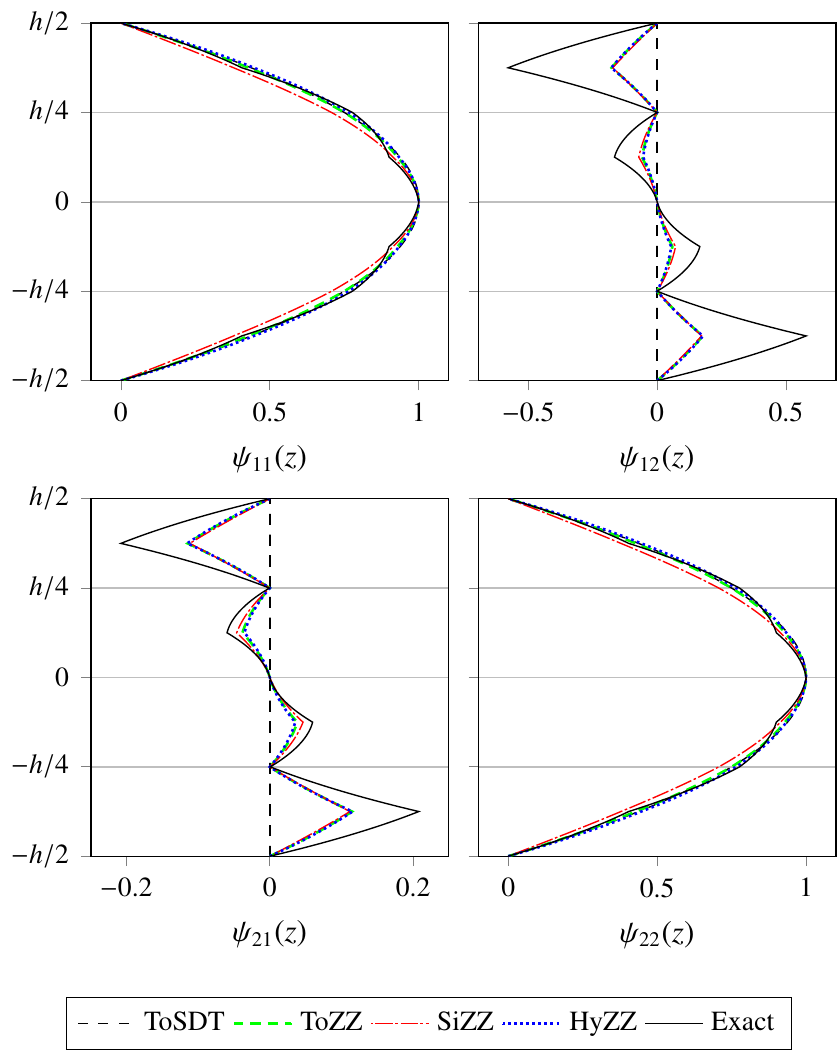}
  }{
    \input{Figures/p30m30x4_stress_func.tex}
  }
  \caption{Transverse shear \sfs{} for the $[30/{-30}]_4$ laminate with $a/h=10$}%
  \label{fig:p30m30x4_psi}
\end{SmartFigure}
\iftoggle{submission}{}{\tikzsetnextfilename{p30m30x4_sig}}
\begin{SmartFigure}[htbp]
  \centering
  \iftoggle{submission}{
    \includegraphics{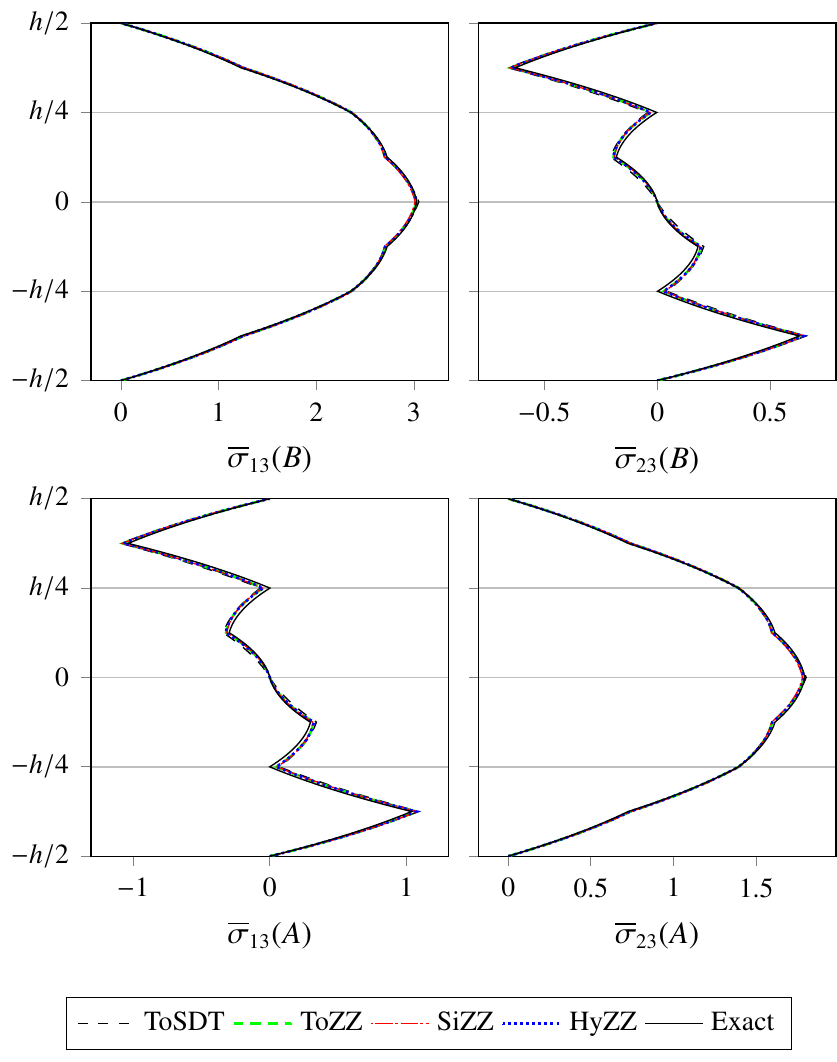}
  }{
    \input{Figures/p30m30x4_sig.tex}
  }
  \caption{Transverse shear stresses from equilibrium equations for the $[30/{-30}]_4$ laminate with $a/h=10$}%
  \label{fig:p30m30x4_sig}
\end{SmartFigure}
%
\begin{table*}[htbp]%
	\centering
	\footnotesize
  \setlength{\tabcolsep}{3pt}
	\begin{tabular}{c@{~}l|c@{~}c|l@{~}c|c@{~}c|c@{~}c|c@{~}c|c@{~}c|c@{~}c|c@{~}c}
    \iftoggle{submission}{
      Seq. & Model & $\overline{w}(C)$ & \% & $\overline{\sigma}_{13}(B)$ & \% & $\overline{\sigma}_{23}(A)$ & \% & $\overline{\sigma}_{23}(B')$ & \% & $\overline{\sigma}_{13}(A')$ & \% & $\overline{\sigma}_{11}(C^+)$ & \% & $\overline{\sigma}_{22}(C^+)$ & \% & $\overline{\omega}$ & \% \\ 

      \hline
      $[30/{-30}]$ & Sin         & $0.83228$ & $-2.86$ & $1.4199$ & $-5.85$ & $0.79419$ & $-7.31$ & $-1.0567$ & $+5.46$ & $-1.7445$ & $+6.44$ & $4.2767$ & $+1.73$ & $1.6979$ & $+0.13$ & $10.864$ & $+1.62$ \\ 
           & \sizzfour{} & $0.82045$ & $-4.24$ & $1.3835$ & $-8.27$ & $0.78021$ & $-8.94$ & $-1.0770$ & $+7.49$ & $-1.7768$ & $+8.41$ & $4.2448$ & $+0.97$ & $1.6858$ & $-0.58$ & $10.943$ & $+2.36$ \\ 
           & Exact       & $0.85678$ &  & $1.5081$ &  & $0.85683$ &  & $-1.0020$ &  & $-1.6390$ &  & $4.2042$ &  & $1.6956$ &  & $10.691$ &  \\ 

      \hline
      $[30/{-30}]_2$ & Sin         & $0.49527$ & $-11.9$ & $3.4961$ & $+3.39$ & $2.0650$ & $+3.50$ & $-0.94454$ & $+7.18$ & $-1.5625$ & $+7.75$ & $2.7054$ & $-4.49$ & $1.0453$ & $-6.23$ & $14.151$ & $+6.63$ \\ 
             & \sizzfour{} & $0.55293$ & $-1.60$ & $3.4038$ & $+0.66$ & $2.0066$ & $+0.58$ & $-0.90016$ & $+2.14$ & $-1.4912$ & $+2.84$ & $2.8315$ & $-0.03$ & $1.0966$ & $-1.63$ & $13.397$ & $+0.95$ \\ 
             & Exact       & $0.56190$ &  & $3.3814$ &  & $1.9950$ &  & $-0.88129$ &  & $-1.4501$ &  & $2.8325$ &  & $1.1147$ &  & $13.270$ &  \\ 
      \hline
      $[30/{-30}]_3$ & Sin         & $0.46217$ & $-9.14$ & $2.8292$ & $-1.76$ & $1.6702$ & $-1.99$ & $-0.37059$ & $+22.4$ & $-0.61259$ & $+23.3$ & $2.6441$ & $-0.67$ & $1.0140$ & $-2.22$ & $14.658$ & $+5.00$ \\ 
             & \sizzfour{} & $0.50452$ & $-0.82$ & $2.8254$ & $-1.89$ & $1.6669$ & $-2.18$ & $-0.33604$ & $+10.9$ & $-0.55759$ & $+12.2$ & $2.6963$ & $+1.29$ & $1.0366$ & $-0.04$ & $14.034$ & $+0.54$ \\ 
             & Exact       & $0.50868$ &  & $2.8800$ &  & $1.7040$ &  & $-0.30288$ &  & $-0.49695$ &  & $2.6619$ &  & $1.0370$ &  & $13.959$ &  \\ 

      \hline
      $[30/{-30}]_4$ & Sin         & $0.45156$ & $-7.89$ & $3.0408$ & $+0.43$ & $1.7972$ & $+0.20$ & $-0.04576$ & n.c. & $-0.07499$ & n.c. & $2.6519$ & $+0.59$ & $1.0135$ & $-0.80$ & $14.832$ & $+4.28$ \\ 
             & \sizzfour{} & $0.48798$ & $-0.46$ & $3.0121$ & $-0.52$ & $1.7806$ & $-0.72$ & $-0.03260$ & n.c. & $-0.05420$ & n.c. & $2.6851$ & $+1.85$ & $1.0284$ & $+0.66$ & $14.273$ & $+0.35$ \\ 
             & Exact       & $0.49024$ &  & $3.0279$ &  & $1.7936$ &  & $-0.00187$ &  & $0.00063$ &  & $2.6364$ &  & $1.0217$ &  & $14.223$ &  \\
      \hline
      $[30/{-30}]_5$ & Sin         & $0.44680$ & $-7.25$ & $2.8815$ & $-0.75$ & $1.7026$ & $-1.05$ & $-0.16727$ & $+20.0$ & $-0.27647$ & $+21.4$ & $2.6677$ & $+1.00$ & $1.0174$ & $-0.29$ & $14.912$ & $+3.92$ \\ 
             & \sizzfour{} & $0.48045$ & $-0.27$ & $2.8700$ & $-1.15$ & $1.6967$ & $-1.39$ & $-0.15842$ & $+13.7$ & $-0.26252$ & $+15.3$ & $2.6952$ & $+2.04$ & $1.0300$ & $+0.95$ & $14.385$ & $+0.25$ \\ 
             & Exact       & $0.48174$ &  & $2.9033$ &  & $1.7206$ &  & $-0.13937$ &  & $-0.22772$ &  & $2.6413$ &  & $1.0203$ &  & $14.349$ &  \\ 

      \hline
      $[30/{-30}]_{10}$ & Sin         & $0.44060$ & $-6.34$ & $2.9297$ & $+0.00$ & $1.7318$ & $-0.32$ & $-0.16383$ & $+11.7$ & $-0.27105$ & $+12.2$ & $2.7238$ & $+0.90$ & $1.0347$ & $-0.15$ & $15.018$ & $+3.41$ \\ 
                & \sizzfour{} & $0.47056$ & $+0.03$ & $2.9115$ & $-0.62$ & $1.7227$ & $-0.84$ & $-0.15541$ & $+5.95$ & $-0.25769$ & $+6.69$ & $2.7522$ & $+1.95$ & $1.0471$ & $+1.04$ & $14.537$ & $+0.10$ \\ 
                & Exact       & $0.47043$ &  & $2.9296$ &  & $1.7374$ &  & $-0.14667$ &  & $-0.24154$ &  & $2.6996$ &  & $1.0363$ &  & $14.523$ &  \\ 

   }{
      
      \hline
      
      \hline
      
      \hline
      
      \hline
      
      \hline
      
      \hline
      
   }  
	\end{tabular}
	\normalsize
	\caption{Comparison between the Sin and the \protect\sizzfour{} models for the $[30/{-30}]_n$ with $n=1,2,3,4,5,10$ ($a/h=10$)}
	\label{tab:p30m30n}
\end{table*}
%
\begin{table*}[htbp]%
	\centering
	\footnotesize
  \setlength{\tabcolsep}{3pt}
	\begin{tabular}{c@{~}l|c@{~}c|l@{~}c|c@{~}c|c@{~}c|c@{~}c|c@{~}c|c@{~}c|c@{~}c}
    \iftoggle{submission}{
      a/h & Model & $\overline{w}(C)$ & \% & $\overline{\sigma}_{13}(B)$ & \% & $\overline{\sigma}_{23}(A)$ & \% & $\overline{\sigma}_{23}(B')$ & \% & $\overline{\sigma}_{13}(A')$ & \% & $\overline{\sigma}_{11}(C^+)$ & \% & $\overline{\sigma}_{22}(C^+)$ & \% & $\overline{\omega}$ & \% \\ 

      \hline
      4 & ToSDT        & $1.4047$ & $-16.8$ & $3.2615$ & $+17.1$ & $1.9714$ & $+16.1$ & $-0.88282$ & $+45.9$ & $-1.4801$ & $+49.3$ & $4.0098$ & $-8.91$ & $1.5636$ & $-14.6$ & $8.3994$ & $+10.8$ \\ 
  & \tozzfour{}  & $1.6454$ & $-2.49$ & $2.7691$ & $-0.57$ & $1.6823$ & $-0.94$ & $-0.65489$ & $+8.26$ & $-1.1232$ & $+13.3$ & $4.5798$ & $+4.04$ & $1.8028$ & $-1.56$ & $7.7493$ & $+2.20$ \\ 
  & Sin          & $1.4016$ & $-16.9$ & $3.2354$ & $+16.2$ & $1.9561$ & $+15.2$ & $-0.87575$ & $+44.8$ & $-1.4684$ & $+48.1$ & $4.1035$ & $-6.78$ & $1.5997$ & $-12.6$ & $8.4065$ & $+10.9$ \\ 
  & \sizzfour{}  & $1.6509$ & $-2.16$ & $2.7252$ & $-2.14$ & $1.6531$ & $-2.66$ & $-0.62965$ & $+4.09$ & $-1.0825$ & $+9.21$ & $4.6690$ & $+6.07$ & $1.8418$ & $+0.58$ & $7.7311$ & $+1.96$ \\ 
  & Hyp          & $1.4049$ & $-16.7$ & $3.2709$ & $+17.4$ & $1.9769$ & $+16.4$ & $-0.88534$ & $+46.3$ & $-1.4842$ & $+49.7$ & $3.9752$ & $-9.70$ & $1.5503$ & $-15.3$ & $8.3994$ & $+10.9$ \\ 
  & \hyzzfour{}  & $1.6417$ & $-2.71$ & $2.7856$ & $+0.02$ & $1.6931$ & $-0.30$ & $-0.66422$ & $+9.80$ & $-1.1382$ & $+14.8$ & $4.5446$ & $+3.24$ & $1.7875$ & $-2.39$ & $7.7597$ & $+2.34$ \\ 
  & Exact        & $1.6875$ &  & $2.7849$ &  & $1.6982$ &  & $-0.60493$ &  & $-0.99119$ &  & $4.4020$ &  & $1.8313$ &  & $7.5824$ &  \\ 

      \hline
      10 & ToSDT       & $0.49462$ & $-12.0$ & $3.5008$ & $+3.53$ & $2.0676$ & $+3.64$ & $-0.94579$ & $+7.32$ & $-1.5645$ & $+7.89$ & $2.6881$ & $-5.10$ & $1.0386$ & $-6.83$ & $14.160$ & $+6.71$ \\ 
   & \tozzfour{} & $0.54918$ & $-2.26$ & $3.4129$ & $+0.93$ & $2.0131$ & $+0.90$ & $-0.90522$ & $+2.71$ & $-1.4996$ & $+3.41$ & $2.8104$ & $-0.78$ & $1.0878$ & $-2.41$ & $13.443$ & $+1.30$ \\ 
   & Sin         & $0.49527$ & $-11.9$ & $3.4961$ & $+3.39$ & $2.0650$ & $+3.50$ & $-0.94454$ & $+7.18$ & $-1.5625$ & $+7.75$ & $2.7054$ & $-4.49$ & $1.0453$ & $-6.23$ & $14.151$ & $+6.63$ \\ 
   & \sizzfour{} & $0.55293$ & $-1.60$ & $3.4038$ & $+0.66$ & $2.0066$ & $+0.58$ & $-0.90016$ & $+2.14$ & $-1.4912$ & $+2.84$ & $2.8315$ & $-0.03$ & $1.0966$ & $-1.63$ & $13.397$ & $+0.95$ \\ 
   & Hyp         & $0.49428$ & $-12.0$ & $3.5024$ & $+3.58$ & $2.0686$ & $+3.69$ & $-0.94624$ & $+7.37$ & $-1.5652$ & $+7.94$ & $2.6818$ & $-5.32$ & $1.0362$ & $-7.04$ & $14.165$ & $+6.74$ \\ 
   & \hyzzfour{} & $0.54759$ & $-2.55$ & $3.4162$ & $+1.03$ & $2.0154$ & $+1.02$ & $-0.90704$ & $+2.92$ & $-1.5026$ & $+3.62$ & $2.8025$ & $-1.06$ & $1.0845$ & $-2.71$ & $13.463$ & $+1.45$ \\ 
   & Exact       & $0.56190$ &  & $3.3814$ &  & $1.9950$ &  & $-0.88129$ &  & $-1.4501$ &  & $2.8325$ &  & $1.1147$ &  & $13.270$ &  \\ 

      \hline
      100 & ToSDT        & $0.31623$ & $-0.23$ & $3.5499$ & $+0.04$ & $2.0842$ & $+0.04$ & $-0.95861$ & $+0.07$ & $-1.5803$ & $+0.08$ & $2.4333$ & $-0.07$ & $0.93570$ & $-0.09$ & $17.781$ & $+0.12$ \\ 
    & \tozzfour{}  & $0.31681$ & $-0.05$ & $3.5490$ & $+0.01$ & $2.0837$ & $+0.01$ & $-0.95819$ & $+0.03$ & $-1.5796$ & $+0.04$ & $2.4346$ & $-0.02$ & $0.93622$ & $-0.04$ & $17.765$ & $+0.03$ \\ 
    & Sin          & $0.31624$ & $-0.23$ & $3.5498$ & $+0.04$ & $2.0842$ & $+0.04$ & $-0.95860$ & $+0.07$ & $-1.5803$ & $+0.08$ & $2.4335$ & $-0.06$ & $0.93577$ & $-0.08$ & $17.781$ & $+0.12$ \\ 
    & \sizzfour{}  & $0.31685$ & $-0.04$ & $3.5489$ & $+0.01$ & $2.0836$ & $+0.01$ & $-0.95814$ & $+0.02$ & $-1.5795$ & $+0.03$ & $2.4348$ & $-0.01$ & $0.93631$ & $-0.03$ & $17.764$ & $+0.02$ \\ 
    & Hyp          & $0.31623$ & $-0.23$ & $3.5499$ & $+0.04$ & $2.0842$ & $+0.04$ & $-0.95861$ & $+0.07$ & $-1.5803$ & $+0.08$ & $2.4332$ & $-0.07$ & $0.93568$ & $-0.09$ & $17.781$ & $+0.12$ \\ 
    & \hyzzfour{}  & $0.31679$ & $-0.06$ & $3.5490$ & $+0.01$ & $2.0837$ & $+0.01$ & $-0.95821$ & $+0.03$ & $-1.5796$ & $+0.04$ & $2.4345$ & $-0.02$ & $0.93618$ & $-0.04$ & $17.765$ & $+0.03$ \\ 
    & Exact        & $0.31697$ &  & $3.5486$ &  & $2.0834$ &  & $-0.95791$ &  & $-1.5790$ &  & $2.4350$ &  & $0.93656$ &  & $17.760$ &  \\ 

   }{
      
      \hline
      
      \hline
      
      \hline
      
   }  
  \end{tabular}
	\normalsize
	\caption{Comparison between the different models for the square $[30/{-30}]_2$ plate with a varying length to thickness ratio}
	\label{tab:p30m30x2}
\end{table*}%
%

%

%
%
%
\subsection{A general non-symmetric anisotropic laminate}
In order to give evidence of the generality of the proposed method for constructing an mHSDT, a laminate with a stacking sequence of the most general nature is next considered. For this example, the arbitrary stacking sequence $[-20/40/70/{-15}/0/40/{-60}]$ has been taken. The four \wfs{} of the polynomial, trigonometric and hyperbolic mHSDT are plotted in figure~\ref{fig:m20p40p70m10p0p40m60_phi} along with the \wfs{} of the polynomial \tosdt{} and of the 3D solution. The corresponding \sfs{} are plotted in figure~\ref{fig:m20p40p70m10p0p40m60_psi}. The distribution across the thickness at points $A$ and $B$ of the transverse shear stresses obtained from the equilibrium equations are illustrated in figure~\ref{fig:m20p40p70m10p0p40m60_sig}. As in the previous examples, all curves are in good agreement. 
\par
Numerical results for the global and local stress response are reported in table~\ref{tab:m20p40p70m10p0p40m60}. A substantial improvement is evident of the predictions provided by the mHSDT over those provided by the bHSDT for laminates with low length-to-thickness ratios, i.e., when the transverse shear behavior plays a certain role. The improvement is here clearly visible not only for the global response (deflection and fundamental frequency), but also for the local bending and transverse shear stresses.
\iftoggle{submission}{}{\tikzsetnextfilename{m20p40p70m10p0p40m60_10_phi}}
\begin{SmartFigure}[htbp]
  \centering
  \iftoggle{submission}{
    \includegraphics{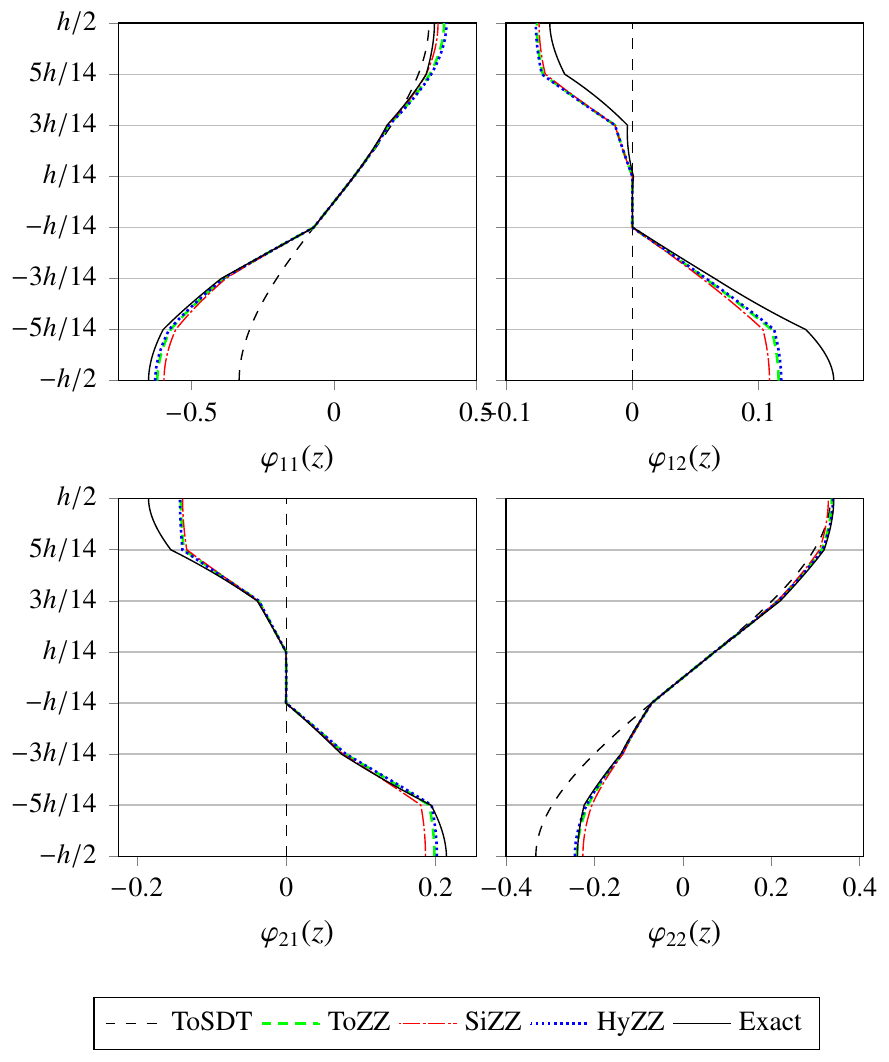}
  }{
    \input{Figures/m20p40p70m10p0p40m60_10_phi.tex}%
   }%
  \caption{\Wfs{} of the $[-20/40/70/{-15}/0/40/{-60}]$ laminate with $a/h=10$ for each considered model}%
  \label{fig:m20p40p70m10p0p40m60_phi}
\end{SmartFigure}
\iftoggle{submission}{}{\tikzsetnextfilename{m20p40p70m10p0p40m60_stress_func}}
\begin{SmartFigure}[htbp]
  \centering
  \iftoggle{submission}{
    \includegraphics{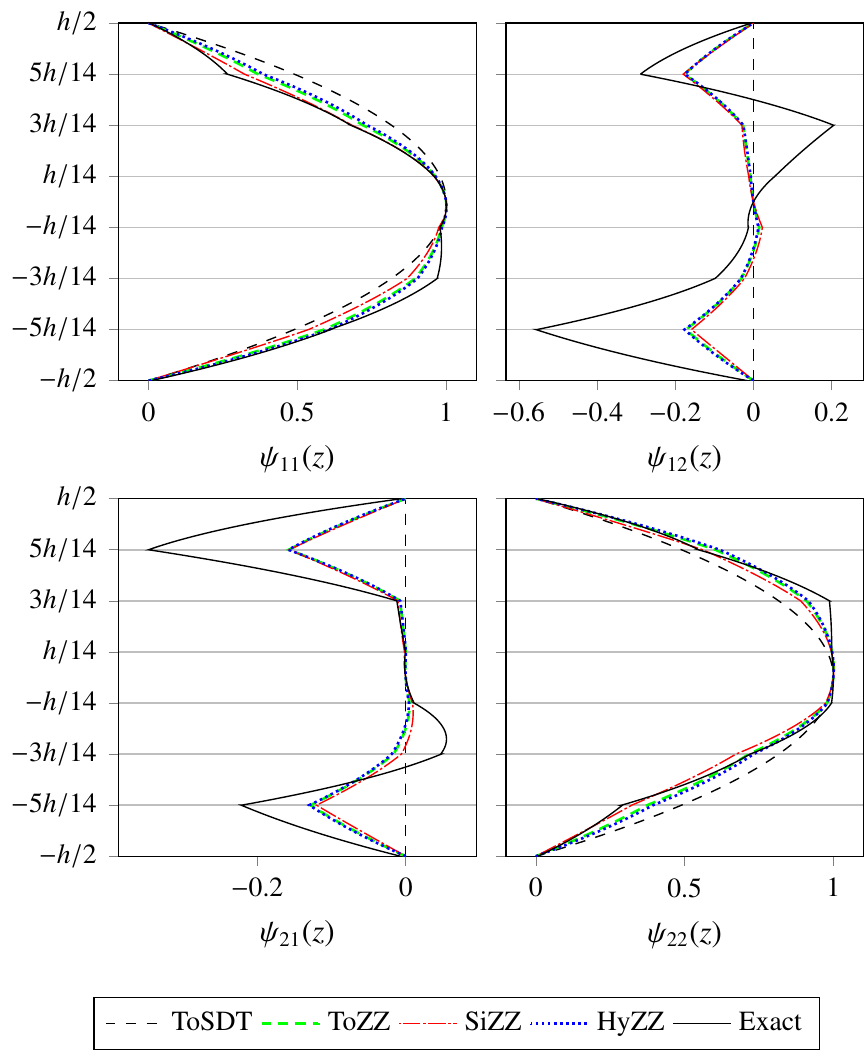}
  }{
    \input{Figures/m20p40p70m10p0p40m60_stress_func.tex}%
  }%
  \caption{Transverse shear \sfs{} of the $[-20/40/70/{-15}/0/40/{-60}]$ laminate with $a/h=10$ for each considered model}%
  \label{fig:m20p40p70m10p0p40m60_psi}
\end{SmartFigure}
\iftoggle{submission}{}{\tikzsetnextfilename{m20p40p70m10p0p40m60_sig}}
\begin{SmartFigure}[htbp]
  \centering
  \iftoggle{submission}{
    \includegraphics{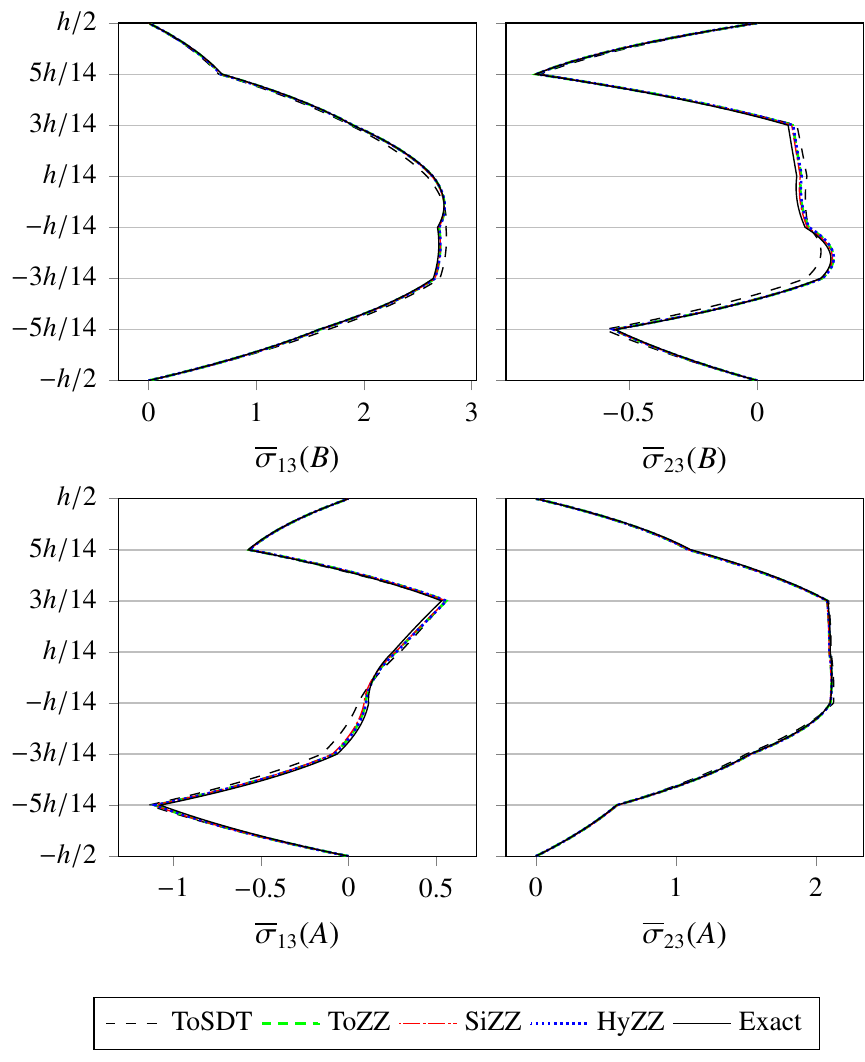}
  }{
    \input{Figures/m20p40p70m10p0p40m60_sig.tex}%
   }%
  \caption{Post-processed transverse shear stresses of the $[-20/40/70/{-15}/0/40/{-60}]$ laminate with $a/h=10$ for each considered model}%
  \label{fig:m20p40p70m10p0p40m60_sig}
\end{SmartFigure}
%
\begin{table*}[htbp]%
	\centering
	\footnotesize
  \setlength{\tabcolsep}{3pt}
	\begin{tabular}{c@{~}l|c@{~}c|l@{~}c|c@{~}c|c@{~}c|c@{~}c|c@{~}c|c@{~}c|c@{~}c}
    \iftoggle{submission}{
      a/h & Model & $\overline{w}(C)$ & \% & $\overline{\sigma}_{13}(B)$ & \% & $\overline{\sigma}_{23}(A)$ & \% & $\overline{\sigma}_{23}(B)$ & \% & $\overline{\sigma}_{13}(A)$ & \% & $\overline{\sigma}_{11}(C^+)$ & \% & $\overline{\sigma}_{22}(C^+)$ & \% & $\overline{\omega}$ & \% \\ 

      \hline
      4 & ToSDT       & $1.4264$ & $-14.3$ & $2.6031$ & $-3.81$ & $2.0206$ & $+3.95$ & $0.24342$ & $+121$  & $0.28916$ & $+137$  & $1.3530$ & $-14.6$ & $3.2716$ & $-5.03$ & $8.3309$ & $+8.83$ \\ 
  & \tozzfour{} & $1.6580$ & $-0.35$ & $2.6973$ & $-0.34$ & $1.9162$ & $-1.42$ & $0.11628$ & $+5.77$ & $0.13017$ & $+6.54$ & $1.4948$ & $-5.61$ & $3.3506$ & $-2.73$ & $7.7204$ & $+0.86$ \\ 
  & Sin         & $1.4220$ & $-14.5$ & $2.5788$ & $-4.71$ & $1.9956$ & $+2.67$ & $0.20552$ & $+86.9$ & $0.25560$ & $+109$  & $1.3788$ & $-12.9$ & $3.3185$ & $-3.66$ & $8.3416$ & $+8.97$ \\ 
  & \sizzfour{} & $1.6558$ & $-0.48$ & $2.6733$ & $-1.22$ & $1.9056$ & $-1.96$ & $0.11431$ & $+3.98$ & $0.12505$ & $+2.35$ & $1.5073$ & $-4.83$ & $3.3578$ & $-2.52$ & $7.7234$ & $+0.90$ \\ 
  & Hyp         & $1.4270$ & $-14.2$ & $2.6116$ & $-3.50$ & $2.0293$ & $+4.40$ & $0.25631$ & $+133$  & $0.30045$ & $+146$  & $1.3434$ & $-15.2$ & $3.2537$ & $-5.55$ & $8.3299$ & $+8.82$ \\ 
  & \hyzzfour{} & $1.6581$ & $-0.34$ & $2.7057$ & $-0.02$ & $1.9200$ & $-1.22$ & $0.11780$ & $+7.15$ & $0.13219$ & $+8.20$ & $1.4892$ & $-5.97$ & $3.3483$ & $-2.80$ & $7.7210$ & $+0.87$ \\ 
  & Exact       & $1.6637$ &  & $2.7063$ &  & $1.9437$ &  & $0.10994$ &  & $0.12217$ &  & $1.5837$ &  & $3.4447$ &  & $7.6547$ &  \\ 

      \hline
      10 & ToSDT       & $0.49707$ & $-8.91$ & $2.7296$ & $-0.43$ & $2.1253$ & $+0.53$ & $0.18717$ & $+20.2$ & $0.15333$ & $+6.56$ & $0.94788$ & $-5.21$ & $2.3218$ & $-2.19$ & $14.113$ & $+4.85$ \\ 
   & \tozzfour{} & $0.54300$ & $-0.49$ & $2.7445$ & $+0.11$ & $2.1111$ & $-0.15$ & $0.17145$ & $+10.1$ & $0.14648$ & $+1.80$ & $0.97417$ & $-2.58$ & $2.3424$ & $-1.32$ & $13.510$ & $+0.37$ \\ 
   & Sin         & $0.49678$ & $-8.96$ & $2.7247$ & $-0.61$ & $2.1191$ & $+0.23$ & $0.17065$ & $+9.58$ & $0.13580$ & $-5.62$ & $0.95372$ & $-4.62$ & $2.3332$ & $-1.71$ & $14.117$ & $+4.88$ \\ 
   & \sizzfour{} & $0.54313$ & $-0.47$ & $2.7401$ & $-0.05$ & $2.1081$ & $-0.29$ & $0.16730$ & $+7.42$ & $0.14065$ & $-2.25$ & $0.97699$ & $-2.30$ & $2.3450$ & $-1.21$ & $13.508$ & $+0.36$ \\ 
   & Hyp         & $0.49704$ & $-8.92$ & $2.7313$ & $-0.37$ & $2.1275$ & $+0.63$ & $0.19272$ & $+23.7$ & $0.15920$ & $+10.6$ & $0.94577$ & $-5.42$ & $2.3177$ & $-2.36$ & $14.114$ & $+4.85$ \\ 
   & \hyzzfour{} & $0.54278$ & $-0.53$ & $2.7462$ & $+0.17$ & $2.1120$ & $-0.11$ & $0.17309$ & $+11.1$ & $0.14837$ & $+3.11$ & $0.97289$ & $-2.71$ & $2.3412$ & $-1.37$ & $13.513$ & $+0.39$ \\ 
   & Exact       & $0.54569$ &  & $2.7414$ &  & $2.1142$ &  & $0.15574$ &  & $0.14389$ &  & $0.99995$ &  & $2.3738$ &  & $13.460$ &  \\ 

      \hline
      100 & ToSDT       & $0.31771$ & $-0.16$ & $2.7434$ & $-0.00$ & $2.1414$ & $+0.00$ & $0.13086$ & $+0.17$ & $0.063505$ & $-0.32$ & $0.87950$ & $-0.06$ & $2.1673$ & $-0.03$ & $17.739$ & $+0.08$ \\ 
    & \tozzfour{} & $0.31819$ & $-0.01$ & $2.7436$ & $+0.00$ & $2.1413$ & $-0.00$ & $0.13087$ & $+0.18$ & $0.063699$ & $-0.02$ & $0.87975$ & $-0.04$ & $2.1675$ & $-0.02$ & $17.726$ & $+0.01$ \\ 
    & Sin         & $0.31771$ & $-0.16$ & $2.7433$ & $-0.01$ & $2.1413$ & $-0.00$ & $0.13061$ & $-0.03$ & $0.063224$ & $-0.76$ & $0.87957$ & $-0.06$ & $2.1675$ & $-0.02$ & $17.739$ & $+0.08$ \\ 
    & \sizzfour{} & $0.31819$ & $-0.01$ & $2.7435$ & $+0.00$ & $2.1412$ & $-0.00$ & $0.13079$ & $+0.11$ & $0.063590$ & $-0.19$ & $0.87979$ & $-0.03$ & $2.1675$ & $-0.02$ & $17.726$ & $+0.01$ \\ 
    & Hyp         & $0.31771$ & $-0.16$ & $2.7434$ & $-0.00$ & $2.1414$ & $+0.00$ & $0.13094$ & $+0.23$ & $0.063599$ & $-0.17$ & $0.87947$ & $-0.07$ & $2.1673$ & $-0.03$ & $17.739$ & $+0.08$ \\ 
    & \hyzzfour{} & $0.31819$ & $-0.01$ & $2.7436$ & $+0.00$ & $2.1413$ & $-0.00$ & $0.13090$ & $+0.20$ & $0.063733$ & $+0.04$ & $0.87974$ & $-0.04$ & $2.1675$ & $-0.02$ & $17.726$ & $+0.01$ \\ 
    & Exact       & $0.31823$ &  & $2.7435$ &  & $2.1413$ &  & $0.13064$ &  & $0.063709$ &  & $0.88006$ &  & $2.1679$ &  & $17.725$ &  \\ 

   }{
      
      \hline
      
      \hline
      
      \hline
      
   }  
  \end{tabular}
	\normalsize
	\caption{Comparison between the different models for the square $[-20/40/70/{-15}/0/40/{-60}]$ plate with a varying length to thickness ratio}
	\label{tab:m20p40p70m10p0p40m60}
\end{table*}
%
%
%
%
\subsection{A sandwich with laminated faces}
A symmetric sandwich plate $[0/45/90/{-45}/c]_s$ is finally considered, which consists of a honeycomb core and two laminated faces. The core has a thickness $32 h/ 40$ and each composite layer in the face laminates has the thickness $h/40$. The material properties are those listed in table~\ref{tab:matprop}. In this case, different interfaces are present, i.e., between plies of different orientations as well as between faces and core. The results for all considered bHSDT and corresponding mHSDT are reported in table~\ref{tab:sandwich_aero}. The improvement introduced by the mHSDT is clearly appreciable for all output variables and is more important for low length-to-thickness ratios (short wavelength). Figure~\ref{fig:sandwich_aero_phi} illustrates the \wfs{} and the corresponding \sfs{} are compared in figure~\ref{fig:sandwich_aero_psi}. As in the previous case studies involving laminated plates, the transverse shear stresses across the sandwich section obtained from the equilibrium equations are very accurate even for the bHSDT model, as shown in figure~\ref{fig:sandwich_aero_sig}.
\iftoggle{submission}{}{\tikzsetnextfilename{sandwich_aero_phi}}
\begin{SmartFigure}[htbp]
  \centering
  \iftoggle{submission}{
    \includegraphics{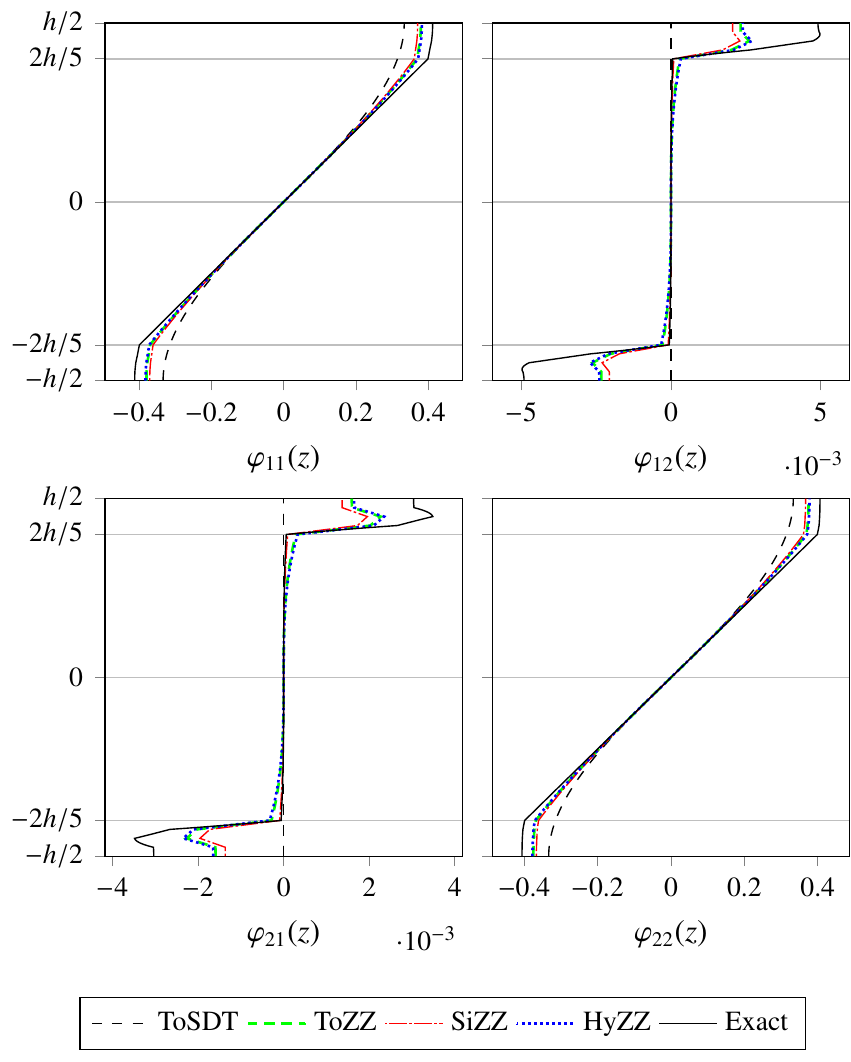}
  }{
    \input{Figures/sandwich_aero_phi.tex}
  }
  \caption{\Wfs{} of the square $[0/45/90/{-45}/c]_s$ sandwich panel with $a/h=10$ for each considered model}%
  \label{fig:sandwich_aero_phi}
\end{SmartFigure}
\iftoggle{submission}{}{\tikzsetnextfilename{sandwich_aero_stress_func}}
\begin{SmartFigure}[htbp]
  \centering
  \iftoggle{submission}{
    \includegraphics{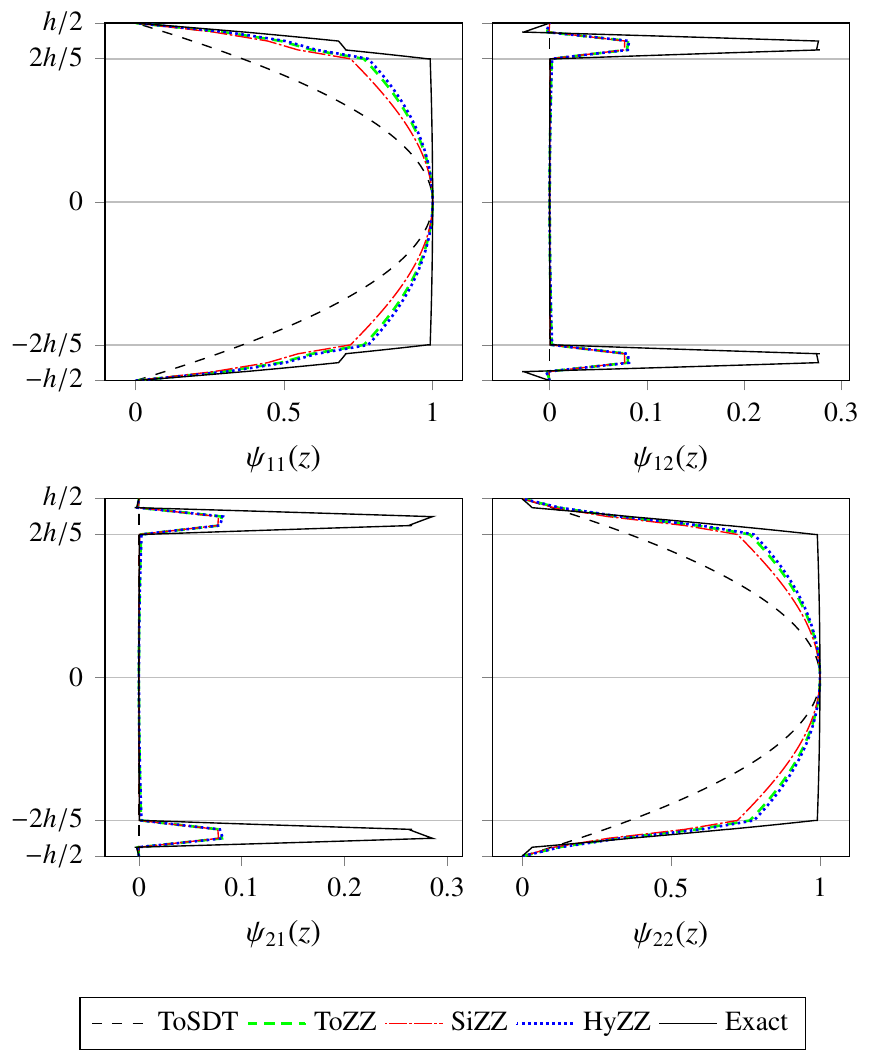}
  }{
    \input{Figures/sandwich_aero_stress_func.tex}
  }
  \caption{Transverse shear \sfs{} of the square $[0/45/90/{-45}/c]_s$ sandwich panel with $a/h=10$ for each considered model}%
  \label{fig:sandwich_aero_psi}
\end{SmartFigure}
\iftoggle{submission}{}{\tikzsetnextfilename{sandwich_aero_sig}}
\begin{SmartFigure}[htbp]
  \centering
  \iftoggle{submission}{
    \includegraphics{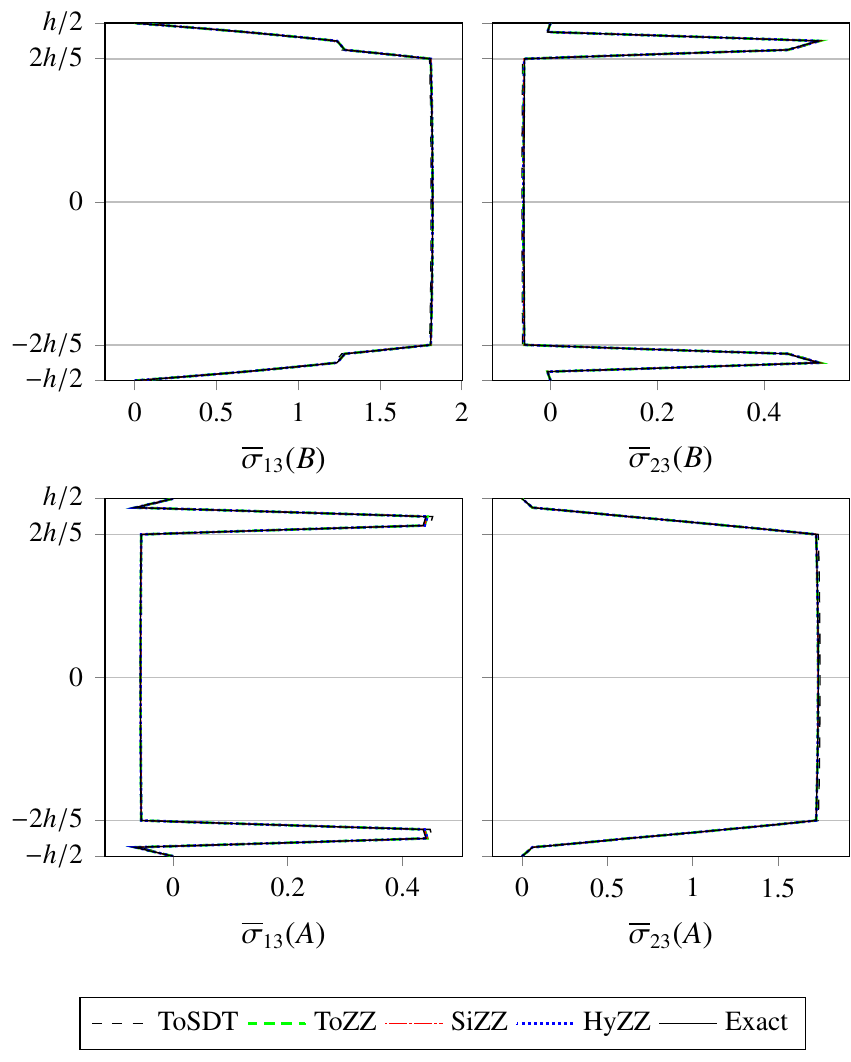}
  }{
    \input{Figures/sandwich_aero_sig.tex}%
  }
  \caption{Post-processed transverse shear stresses of the square $[0/45/90/{-45}/c]_s$ sandwich panel with $a/h=10$ for each considered model}%
  \label{fig:sandwich_aero_sig}
\end{SmartFigure}
%
%
\begin{table*}[htbp]%
	\centering
	\footnotesize
  \setlength{\tabcolsep}{3pt}
	\begin{tabular}{c@{~}l|c@{~}c|l@{~}c|c@{~}c|c@{~}c|c@{~}c|c@{~}c|c@{~}c|c@{~}c}
    \iftoggle{submission}{
      a/h & Model & $\overline{w}(C)$ & \% & $\overline{\sigma}_{13}(B)$ & \% & $\overline{\sigma}_{23}(A)$ & \% & $\overline{\sigma}_{23}(B)$ & \% & $\overline{\sigma}_{13}(A)$ & \% & $\overline{\sigma}_{11}(C^+)$ & \% & $\overline{\sigma}_{22}(C^+)$ & \% & $\overline{\omega}$ & \% \\[2pt] 

      \hline
      4 & ToSDT        & $0.21866$ & $-7.30$ & $1.7883$ & $-1.10$ & $1.7141$ & $+1.29$ & $-0.044424$ & $+33.2$ & $-0.044383$ & $+8.65$ & $11.264$ & $-7.33$ & $0.71690$ & $-17.5$ & $10.804$ & $+5.26$ \\ 
  & \tozzfour{}  & $0.23595$ & $+0.03$ & $1.8046$ & $-0.20$ & $1.6865$ & $-0.33$ & $-0.034844$ & $+4.44$ & $-0.041229$ & $+0.93$ & $12.231$ & $+0.63$ & $0.79273$ & $-8.82$ & $10.307$ & $+0.42$ \\ 
  & Sin          & $0.21785$ & $-7.65$ & $1.7914$ & $-0.93$ & $1.7053$ & $+0.77$ & $-0.043120$ & $+29.2$ & $-0.044101$ & $+7.96$ & $11.417$ & $-6.06$ & $0.72385$ & $-16.7$ & $10.797$ & $+5.20$ \\ 
  & \sizzfour{}  & $0.23523$ & $-0.28$ & $1.8043$ & $-0.22$ & $1.6851$ & $-0.42$ & $-0.034756$ & $+4.18$ & $-0.041184$ & $+0.82$ & $12.250$ & $+0.79$ & $0.79315$ & $-8.77$ & $10.321$ & $+0.56$ \\ 
  & Hyp          & $0.21867$ & $-7.30$ & $1.7871$ & $-1.17$ & $1.7173$ & $+1.49$ & $-0.044891$ & $+34.6$ & $-0.044511$ & $+8.96$ & $11.203$ & $-7.83$ & $0.71396$ & $-17.9$ & $10.813$ & $+5.35$ \\ 
  & \hyzzfour{}  & $0.23615$ & $+0.11$ & $1.8047$ & $-0.19$ & $1.6870$ & $-0.31$ & $-0.034879$ & $+4.55$ & $-0.041251$ & $+0.98$ & $12.225$ & $+0.58$ & $0.79249$ & $-8.84$ & $10.304$ & $+0.39$ \\ 
  & Exact        & $0.23588$ &  & $1.8082$ &  & $1.6922$ &  & $-0.033362$ &  & $-0.040850$ &  & $12.154$ &  & $0.86938$ &  & $10.264$ &  \\ 

      \hline
      10 & ToSDT        & $0.056624$ & $-5.19$ & $1.8124$ & $-0.49$ & $1.7458$ & $+0.59$ & $-0.053324$ & $+7.10$ & $-0.056255$ & $-1.24$ & $8.5063$ & $-2.60$ & $0.44457$ & $-4.96$ & $21.031$ & $+3.36$ \\ 
   & \tozzfour{}  & $0.059592$ & $-0.22$ & $1.8210$ & $-0.02$ & $1.7348$ & $-0.05$ & $-0.050626$ & $+1.68$ & $-0.056688$ & $-0.48$ & $8.7373$ & $+0.05$ & $0.45502$ & $-2.72$ & $20.375$ & $+0.14$ \\ 
   & Sin          & $0.056527$ & $-5.35$ & $1.8143$ & $-0.39$ & $1.7428$ & $+0.41$ & $-0.052963$ & $+6.38$ & $-0.056327$ & $-1.11$ & $8.5485$ & $-2.11$ & $0.44523$ & $-4.82$ & $21.016$ & $+3.29$ \\ 
   & \sizzfour{}  & $0.059484$ & $-0.40$ & $1.8210$ & $-0.02$ & $1.7344$ & $-0.07$ & $-0.050619$ & $+1.67$ & $-0.056669$ & $-0.51$ & $8.7418$ & $+0.10$ & $0.45506$ & $-2.71$ & $20.391$ & $+0.22$ \\ 
   & Hyp          & $0.056613$ & $-5.21$ & $1.8117$ & $-0.53$ & $1.7469$ & $+0.65$ & $-0.053452$ & $+7.36$ & $-0.056234$ & $-1.28$ & $8.4900$ & $-2.78$ & $0.44427$ & $-5.02$ & $21.045$ & $+3.43$ \\ 
   & \hyzzfour{}  & $0.059620$ & $-0.17$ & $1.8209$ & $-0.03$ & $1.7349$ & $-0.04$ & $-0.050629$ & $+1.69$ & $-0.056695$ & $-0.47$ & $8.7358$ & $+0.03$ & $0.45498$ & $-2.73$ & $20.371$ & $+0.12$ \\ 
   & Exact        & $0.059723$ &  & $1.8214$ &  & $1.7356$ &  & $-0.049788$ &  & $-0.056962$ &  & $8.7331$ &  & $0.46776$ &  & $20.347$ &  \\ 

      \hline
      100 & ToSDT        & $0.025743$ & $-0.12$ & $1.8143$ & $-0.01$ & $1.7549$ & $+0.01$ & $-0.054873$ & $+0.09$ & $-0.058710$ & $-0.03$ & $7.9474$ & $-0.04$ & $0.39366$ & $-0.06$ & $30.985$ & $+0.08$ \\ 
    & \tozzfour{}  & $0.025773$ & $-0.01$ & $1.8144$ & $-0.00$ & $1.7547$ & $-0.00$ & $-0.054837$ & $+0.02$ & $-0.058722$ & $-0.01$ & $7.9503$ & $+0.00$ & $0.39375$ & $-0.03$ & $30.962$ & $+0.00$ \\ 
    & Sin          & $0.025742$ & $-0.13$ & $1.8143$ & $-0.01$ & $1.7549$ & $+0.01$ & $-0.054868$ & $+0.08$ & $-0.058712$ & $-0.03$ & $7.9480$ & $-0.03$ & $0.39366$ & $-0.05$ & $30.984$ & $+0.07$ \\ 
    & \sizzfour{}  & $0.025772$ & $-0.01$ & $1.8144$ & $-0.00$ & $1.7547$ & $-0.00$ & $-0.054837$ & $+0.02$ & $-0.058721$ & $-0.01$ & $7.9503$ & $+0.00$ & $0.39375$ & $-0.03$ & $30.963$ & $+0.00$ \\ 
    & Hyp          & $0.025742$ & $-0.12$ & $1.8143$ & $-0.01$ & $1.7549$ & $+0.01$ & $-0.054874$ & $+0.09$ & $-0.058710$ & $-0.03$ & $7.9472$ & $-0.04$ & $0.39366$ & $-0.06$ & $30.985$ & $+0.08$ \\ 
    & \hyzzfour{}  & $0.025773$ & $-0.01$ & $1.8144$ & $-0.00$ & $1.7547$ & $-0.00$ & $-0.054837$ & $+0.02$ & $-0.058722$ & $-0.01$ & $7.9503$ & $+0.00$ & $0.39375$ & $-0.03$ & $30.962$ & $+0.00$ \\ 
    & Exact        & $0.025774$ &  & $1.8144$ &  & $1.7548$ &  & $-0.054825$ &  & $-0.058728$ &  & $7.9503$ &  & $0.39387$ &  & $30.962$ &  \\ 

   }{
      
      \hline
      
      \hline
      
      \hline
      
   }  
  \end{tabular}
	\normalsize
	\caption{Comparison between the different models for the square $[0/45/90/{-45}/c]_s$ sandwich panel with a varying length to thickness ratio}
	\label{tab:sandwich_aero}
\end{table*}%
%
%
%
%
\section{Discussion: limitations of ZZ theories}\label{sec:discussion}
The presented results show that the mHSDT, constructed in such a manner that all $C_z^0-$requirements are {\em a priori} satisfied, can provide very accurate solutions in terms of deflection, transverse stresses and fundamental frequency for a very large class of composite plates. In particular, mHSDT improve the results of bHSDT for moderately thick plates that consist of a quite large number of layers. 

As a matter of fact, for configurations with a very low number of layers and strong anisotropy, the constructed mHSDT appear to be not better than the corresponding bHSDT. This is true in general for any ZZ theory and is due to the ESL nature of these theories. In order to point out the limitations of these ZZ theories, we shall next consider the \wfs{} obtained from the exact 3D solution \cite{Loredo2014,Loredo2016}. While such \wfs{} can be defined for general laminates only in a numerical manner, an analytical expression can be found for the elementary case of an orthotropic single-layer plate. Exact solutions obtained, e.g., by a state-space approach are expressed in terms of hyperbolic functions. Therefore, the following two new hyperbolic \wfs{} are proposed:

\begin{equation}\label{eq:Hyp2_wf}
  \left\{
  \begin{array}{l} 
    \displaystyle\varphi_{11}(z)= \frac{1}{a_1-1}\left(a_1 z -\frac{1}{a_2}\sinh{\left(a_2 \frac{z}{h}\right)}\right)\\ 
    \displaystyle\varphi_{22}(z)= \frac{1}{a_3-1}\left(a_3 z -\frac{1}{a_4}\sinh{\left(a_4 \frac{z}{h}\right)}\right)
  \end{array}
  \right. 
\end{equation}
where the coefficients $a_1, a_2, a_3, a_4$ have the expressions
\begin{equation}
  \left\{
  \begin{array}{*3{>{\displaystyle}l}} 
    \displaystyle a_1=\cosh{\left(\frac{a_2}{2}\right)} &;\;& a_2=\frac{\pi h}{l_x}\sqrt{\frac{Q_{1111}}{G_{13}}}\\ 
    \displaystyle a_3=\cosh{\left(\frac{a_4}{2}\right)} &;\;& a_4=\frac{\pi h}{l_y}\sqrt{\frac{Q_{2222}}{G_{23}}}\\ 
  \end{array}
  \right. 
\end{equation}
and where $l_x$ and $l_y$ denote the lengths of the plate edges along the $x$ and $y$ directions, respectively.

It is emphasised that the resulting model, referred to as \hyspe{} model, is not a member of the hyperbolic mHSDT described in Section~\ref{sec:ext}. Two major points should be remarked: first, the \hyspe{} model employs two odd functions instead of only one odd function (see equation~\eqref{eq:hyzz}); second, the material properties of the layer as well as the length-to-thickness ratio $l/h$ appear in the argument of the hyperbolic function. This \hyspe{} model is next compared against the corresponding ``conventional'' ZZ model \hyzzfour{}, whose analytical expression is reported in Appendix, equation~\eqref{eq:hyzz} with $k=2$, in order to point out the limitations of the latter model. 

\iftoggle{submission}{}{\tikzsetnextfilename{OrthoSpecial_stress_func}}
\begin{SmartFigure}[htbp]
  \centering
  \iftoggle{submission}{
    \includegraphics{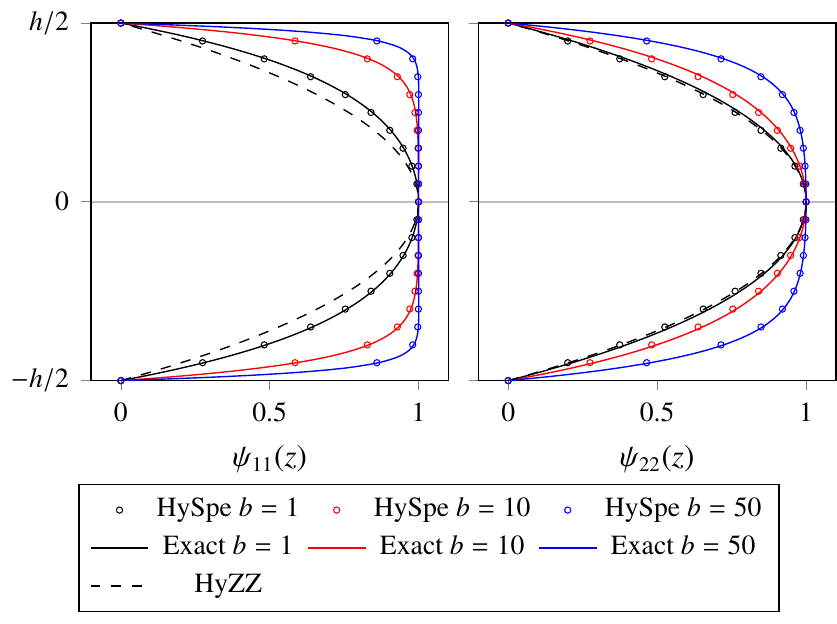}
  }{
    \input{Figures/OrthoSpecial_stress_func.tex}
  }
  \caption{Influence of the shear flexibility factor $b$ on the accuracy of the transverse shear \sfs{} for the \hyzzfour{} and the \hyspe{} models ($[0]$ square plate with $a/h=4$)}%
  \label{fig:OrthoSpecial_psi}
\end{SmartFigure}

Figure~\ref{fig:OrthoSpecial_psi} compares these two hyperbolic HSDT against the exact 3D solution in terms of transverse shear \sfs{} for different values of the elastic moduli: for this, the shear flexibilities $1/G_{23}, 1/G_{13}$ and $1/G_{12}$ of table~\ref{tab:matprop} are multiplied by the same factor $b$. From figure~\ref{fig:OrthoSpecial_psi} it is apparent that the \hyzzfour{} is independent from this coefficient and that it provides a satisfying approximation to the 3D solution only for $b=1$. As the shear flexibility increases, the error of the ``conventional'' \hyzzfour{} model increases while the \sfs{} of the \hyspe{} model are capable of well reproducing the 3D solution.

\iftoggle{submission}{}{\tikzsetnextfilename{p0_stress_func_elanc}}
\begin{SmartFigure}[htbp]
  \centering
  \iftoggle{submission}{
    \includegraphics{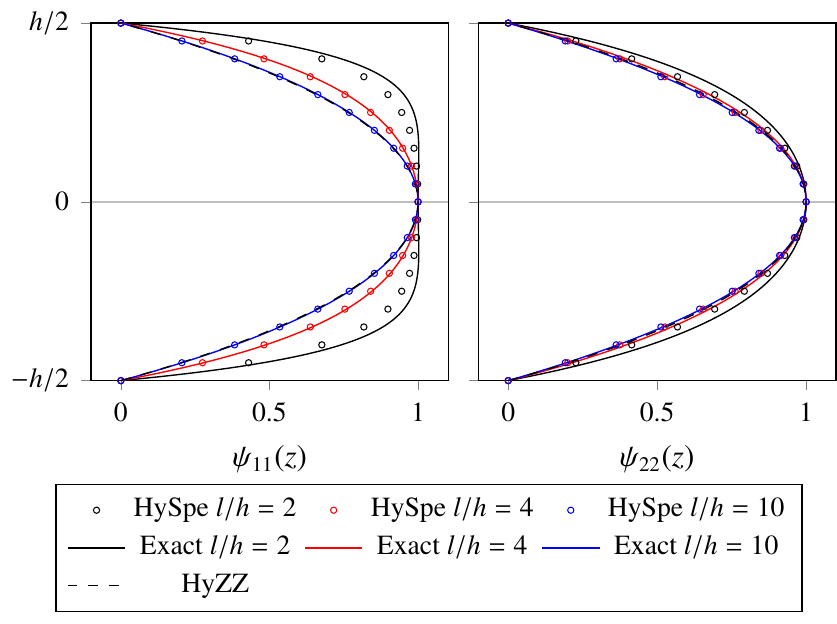}
  }{
    \input{Figures/p0_stress_func_elanc.tex}
  }
  \caption{Influence of the length-to-thickness ratio of the $[0]$ square plate on the accuracy of the transverse shear \sfs{} for the \hyzzfour{} and the \hyspe{} models (orthotropic material of table~\ref{tab:matprop})}%
  \label{fig:p0_stress_func_elanc}
\end{SmartFigure}

The influence of the length-to-thickness ratio on the accuracy of ZZ models is shown in figure~\ref{fig:p0_stress_func_elanc}, where the transverse shear \sfs{} obtained by the \hyspe{} model, the \hyzzfour{} ($k=2$) and the 3D solution are compared for different values of the $l_x/h = l_y/h = a/h$ ratio. The results show that \hyspe{} model has a slight discrepancy with the 3D solution only for the extremely thick case $a/h=2$. Note that this discrepancy comes from the plane stress assumption that has allowed to derive the analytical expressions in equation~\eqref{eq:Hyp2_wf}. Furthermore, the  \hyzzfour{} model recovers the correct solution only if the length-to-thickness ratio is sufficiently large, say $a/h \geq 10$. 

\par

These observations allow to substantiate the limitations of most ``conventional'' ZZ models, such as those that may be formulated within the general procedure proposed in this paper. On the one hand, \wfs{} should depend on the material properties in a far more complex manner than through the linear coefficients $a_{\alpha\beta}$ and $b_{\alpha\beta}$ as identified by equation~\eqref{eq:WarpingFunctions}. More specifically, the \wfs{} should depend on the ratio between the longitudinal and the transverse shear moduli: ZZ models whose \wfs{} depend only on the transverse shear moduli will always suffer a certain limitation with respect to highly constrained configurations such as laminated plates with low number of layers. On the other hand, `conventional'' ZZ models suffer a certain inaccuracy in case of thick laminates unless the length-to-thickness ratio is explicitly taken into account in the \wfs{}.
%
%
%
%
%
\section{Conclusion}\label{sec:conclusion}
In this paper, a general method has been presented to extend basis higher order shear deformation theories (bHSDT) to their multilayer counterpart, denoted mHSDT, which {\em a priori} and exactly meet all $C_z^0-$requirements along with the homogeneous shear stress conditions at the plate's top and bottom surfaces, without introducing any additional DOF. The method is purely displacement-based as no assumption is introduced about the transverse shear stress behaviour. The extension process constructs zig-zag models from a couple of one even and one odd high-order functions of $z$ of arbitrary nature. As examples, \wfs{} based on polynomial, trigonometric and hyperbolic functions have been explicitly addressed by referring to the couples of functions $\left( z^3, z^2 \right)$, $\left( \sin(\pi \frac{z}{h}), \cos(\pi \frac{z}{h}) \right)$, $\left( \sinh(k \frac{z}{h}), \cosh(k \frac{z}{h}) \right)$, respectively. For general stacking sequences of $N$ layers, four \wfs{} are to be determined by solving a linear system of size $8N+8$. Furthermore, the tensorial character of the \wfs{} has been pointed out, which implies that \wfs{} of a stacking sequence $s+\theta$, defined upon a rotation $\theta$ about the $z-$axis of a stacking sequence $s$, must correspond to those of the stacking sequence $s$ through the well known tensorial transformation.
\par
The considered ESL ZZ mHSDT models have been compared against the bHSDT for a large variety of composite plates, including cross-ply, angle-ply, arbitrary laminates and sandwich plates. Exact solutions are obtained based on a previously developed method, for the static bending and the fundamental frequency responses of simply-supported plates. The results indicate that mHSDT substantially increase the accuracy of the bHSDT in most practically relevant applications. No appreciable difference is found between models formulated in terms of polynomial, trigonometric or hyperbolic functions. A final discussion has been proposed in order to point out the limitations of ZZ models with respect to ``constrained'' configurations characterised by a low number of layers (say, $N \le 4$) and low length-to-thickness ratio: in order to recover with good accuracy the exact solution, these configurations require \wfs{} that take into account in an explicit manner also the in-plane stiffnesses of each individual layer. This contrasts with the ``conventional'' ZZ approaches, in which only the transverse shear moduli are considered. While an explicit representation of the mechanical and geometrical properties of each individual layer is inherent to LayerWise models, the possibility of formulating more refined ZZ theories capable of overcoming the limitations of currently available ESL models will be an object of further studies.
\appendix
\section{Analytical expressions for one layer}
For a plate consisting of one layer, the system of equations~\eqref{eq:properties_2} reduces to:
\begin{subequations}
\begin{empheq}[left=\empheqlbrace]{align}
     &d^{1}_{\alpha\beta}=0 \label{eq:prop4a} \\ 
     &a_{\alpha\beta}f'(0^-)+b_{\alpha\beta}g'(0^-)+c^{1}_{\alpha\beta}=\delta_{\alpha\beta} \label{eq:prop4c} \\ 
     &a_{\alpha\beta}f'(-h/2)+b_{\alpha\beta}g'(-h/2)+c^{1}_{\alpha\beta}=0\label{eq:prop4e}\\
     &a_{\alpha\beta}f'(+h/2)+b_{\alpha\beta}g'(+h/2)+c^{1}_{\alpha\beta}=0\label{eq:prop4f}
\end{empheq}\label{eq:properties_4}
\end{subequations}
where $\delta_{\alpha \beta}$ is Kronecker's delta. Since $f'(-h/2)=f'(h/2)$ and $g'(-h/2)=-g'(h/2)$, one has $b_{\alpha\beta}=0$ and $a_{\alpha\beta}f'(h/2)+c^{1}_{\alpha\beta}=0$ and the solution is found to be expressed in terms of only one odd function $f(z)$ as
\begin{equation}
  \varphi_{\alpha\beta}(z)= \frac{\delta_{\alpha\beta}}{f'(0^-)-f'(h/2)}\left(f(z)- z\, f'(h/2) \right)
  \label{eq:phi_one_layer}
\end{equation}
Note that the tensorial character of the \wfs{} is preserved with $\varphi_{11} = \varphi_{22}$; \wfs{} for an arbitrary layer orientation $\theta$ can be obtained through eq.~\eqref{eq:tensTrans}.

For a polynomial HSDT model, setting $f(z)=z^3$ leads to
\begin{equation}
 \varphi_{\alpha\beta}(z) = \delta_{\alpha\beta}\left(z-\frac{4}{3}\frac{z^3}{h^2}\right)
\end{equation}
For a Sinus-model, setting $f(z)=\sin(\pi z/h)$ leads to
\begin{equation}
 \varphi_{\alpha\beta}(z) = \delta_{\alpha\beta}\frac{h}{\pi}\sin\left(\frac{\pi z}{h}\right)
\end{equation}
For a hyperbolic model one may set $f(z)=\sinh(kz/h)$, which yields
\begin{equation}
 \varphi_{\alpha\beta}(z) = \delta_{\alpha\beta}\left(\cosh\left(\frac{k}{2}\right)-1\right)^{-1}\left(\cosh\left(\frac{k}{2}\right)z-\frac{h}{k}\sinh\left(k\frac{z}{h}\right)\right)
\label{eq:hyzz}
\end{equation}
%
%
%
%

%
%
%

%
%
%
%
\label{lastpage}
\end{document}